\documentclass[12pt]{article}
\usepackage{graphicx}
\newfont{\larom}{cmbx10 scaled\magstep3}
\newfont{\bsan}{cmssbx10}
\newenvironment{namelist}[1]{%                             %
\begin{list}{}                                             %
    {                                                      %
      \settowidth{\labelwidth}{#1}                         %
      \setlength{\leftmargin}{1.1\labelwidth}              %          
    }                                                      %
   }{%                                                     %
\end{list}}                                                %
\begin{document}

\begin{center}

  \vspace{8mm}

  {\larom On Modelling a Relativistic Hierarchical (Fractal) Cosmology by 
   Tolman's Spacetime. \\
   III. Numerical Results}  

  \vspace{10mm}
  {\Large Marcelo B. Ribeiro \footnote{ \ Present address: Departamento
  de Astrof\'{\i}sica, Observat\'{o}rio Nacional - CNPq, Rua General
  Jos\'{e} Cristino 77, Rio de Janeiro, RJ 20921, Brazil.} \\}
  \vspace{8mm}
  {\normalsize Astronomy Unit \\ School of Mathematical
  Sciences \\ Queen Mary and Westfield College \\ Mile End Road \\
  London E1 4NS \\ England \\ }

  \vspace{10mm}
% {\bsan Preprint (Revised) \ \ -- \ \ March 1993}
% {\bsan Submitted to the Astrophysical Journal}
% {\bsan to appear in the Astrophysical Journal - 1993}

  \vspace{10mm}
  {\bf ABSTRACT}
\end{center}
  \begin{quotation}
    \small

    This paper presents numerical solutions of particular Tolman models
    approximating a fractal behaviour along the past light cone. The initial
    conditions of the numerical problem are discussed and the algorithm
    used to carry out the numerical integrations is presented. It was
    found that the numerical solutions are stiff across the flat-curved
    interface necessary to obtain the initial conditions of the problem.
    The spatially homogeneous Friedmann models are treated as special
    cases of the Tolman solution and solved numerically. Extending the
    results of paper II on the Einstein-de Sitter model, to the
    $K = \pm 1$ models, it was found that the open and closed Friedmann
    models also do not appear to remain homogeneous along the backward null
    cone, with a vanishing volume (average) density as one approaches the
    big bang singularity hypersurface. Fractal solutions, that is,
    solutions representing an averaged and smoothed-out single fractal,
    were obtained in all three classes of the Tolman metric, but only
    the hyperbolic ones were found to be in agreement with observations,
    meaning that a possible Friedmann background universe would have to be an
    open one. The best fractal metric obtained through numerical
    simulations is also analysed in terms of evolution, homothetic
    self-similarity, comparison with the respective spatially
    homogeneous case and the fitting problem in cosmology. The paper
    finishes with a discussion on some objections raised by some
    authors against a fractal cosmology.

  \end{quotation}
\vspace{6mm}
%\newpage
% \arabic{footnote}
% \setcounter{footnote}{0}

\section{Introduction}
  This is the third paper of a series that studies a
  relativistic cosmology modelling the relativistic generalization
  of the single fractal Newtonian model advanced by
  Pietronero (1987; see also Coleman \& Pietronero 1992), in which the
  galactic clustering problem is studied
  by assuming that the large-scale structure of the universe can be
  described as being a self-similar fractal
  system.\footnote{ \ The models investigated in
  this series of papers are in the realm of classical cosmology. No
  hypotheses concerning inflationary cosmology have been considered.}

  In Ribeiro (1992a, hereafter paper I) I argued that the recent
  all sky redshift surveys (de Lapparent, Geller \& Huchra 1986;
  Saunders et al. 1991) present observations consistent both with the
  old Charlier hypothesis of hierarchical clustering  and with
  fractals, where the latter is in essence a more precise
  conceptualization of the scaling idea implicit in the hierarchical
  clustering hypothesis. In paper I Pietronero's (1987) basic hypotheses
  were assumed in order to propose similar ones in a relativistic context,
  and I obtained
  observational relations compatible with fractals in Tolman's spacetime
  and devised a numerical strategy for finding particular Tolman
  solutions representing a fractal behaviour along the backward null
  cone. In Ribeiro (1992b, hereafter paper II) I studied analytically the
  Einstein-de Sitter model in the context of the theory developed in
  paper I. By treating the Einstein-de Sitter model as a special case of
  Tolman's spacetime, I found that it does not appear to remain
  homogeneous along the past null geodesic, has a volume (average) density which
  vanishes asymptotically and that it also shows no single fractal features
  along the backward null cone. The apparent inhomogeneity of the
  Einstein-de Sitter model is explained by the fact that densities
  measured along the geodesic go through different hypersurfaces of
  constant $t$, where each one has a different value for the proper
  density.

  This paper continues the study of these cosmologies and presents
  relativistic fractal solutions obtained by following the numerical
  simulation strategy already devised in paper I. By {\it fractal
  solutions} I mean solutions where the fractal system is smoothed-out
  and its average density follows the de Vaucouleurs density power-law.
  These solutions represent
  fractal behaviour along the backward null cone and
  they were obtained for all three types of Tolman dust models, namely,
  elliptic, parabolic and hyperbolic. By analysing these solutions we
  conclude that the only ones with features that may represent
  real astronomical observations are of hyperbolic type. As we are
  studying the Tolman spacetime as a region (or regions) possibly surrounded
  by a Friedmann universe (see paper I), if we adopt the fitting condition
  approach for interpreting the match between the two spacetimes
  (Ellis \& Stoeger 1987; Ellis \& Jaklitsch 1989) we then conclude that 
  the Friedmann background required is also hyperbolic and we would
  be living in an open, ever expanding, universe.

  This paper is organized as follows. In \S 2 is shown a summary
  of the observational relations developed in papers I and II plus some minor
  extensions which will be necessary here, and in \S 3 I
  discuss the initial conditions and the algorithm used to find Tolman
  numerical solutions along the past light cone. In \S 4 I present the
  numerical solutions for the spatially homogeneous special cases and
  \S 5 shows fractal solutions for all types
  of Tolman models. \S 6 discusses these fractal solutions in terms of
  comparison with the spatially homogeneous cases, fitting condition, evolution
  of the most realistic fractal model in terms of real observations
  and relations with homothetic self-similarity. In this section I also
  express my criticism of criticisms of fractal cosmology. I finish in
  \S 7 with the conclusions on this and the preceding papers.

\section{Observational Relations in Tolman's Dust~Models}
  The Tolman (1934) metric for the motion of spherically symmetric
  dust is
  \begin{equation}
   dS^2=dt^2-\frac{{R'}^2}{f^2} dr^2 - R^2 (d \theta^2+\sin^2 \theta d \phi^2).
   \label{t1}
  \end{equation}
  The Einstein field equations (with $c=G=1$ and $\Lambda=0$) for metric
  (\ref{t1}) reduce to a single equation
  \begin{equation}
   2 R {\dot{R}}^2 + 2 R (1-f^2) = F,
   \label{t2}
  \end{equation}
  where a dot means $\partial/\partial t$ and a prime means
  $\partial/\partial r$; $f(r)$ and $F(r)$ are two arbitrary functions
  and $R(r,t) \geq 0$. The analytical solutions of equation (\ref{t2})
  are divided into three categories according to the values of the function
  $f(r)$. For $f^2 < 1$ we have the elliptic class and equation
  (\ref{t2}) has the following solution:
  \begin{eqnarray}
      R=\frac{F(1-\cos 2\Theta)}{4 {\mid f^2-1 \mid}},
  \label{u1}
  \end{eqnarray}
  where $\Theta$ is given by
  \begin{eqnarray}
      t+\beta= \frac{F(2\Theta-\sin 2\Theta)}{4{\mid f^2-1 \mid}^{3/2}}.
  \label{u2}
  \end{eqnarray}
  For $f^2 = 1$ we have the parabolic class and the solution
  of equation (\ref{t2}) may be written as
 \begin{eqnarray} 
      R = \frac{1}{2} {(9F)}^{1/3} {(t+\beta)}^{2/3}.
  \label{u3}
 \end{eqnarray}
 Finally the hyperbolic class is given when $f^2 > 1$, and equation
 (\ref{t2}) is solved by
 \begin{eqnarray} 
       R=\frac{F(\cosh 2\Theta-1)}{4(f^2-1)},
   \label{u4}
 \end{eqnarray}
 where
 \begin{eqnarray} 
       t+\beta=\frac{F(\sinh 2\Theta-2\Theta)}{4{(f^2-1)}^{3/2}}.
  \label{u5}
 \end{eqnarray}
 Note that one Tolman solution can have regions of differing classes,
 depending on the values of the function $f(r)$.

 The derivatives $R'$, $\dot{R}$, $\dot{R'}$ for the three cases above
 form long algebraic expressions and I shall not show them here. They
 can be found in paper I. The
 function $\beta (r)$ is a third arbitrary function which gives the
 local time passed since the singularity surface, that is, since the
 big bang. In the elliptic and hyperbolic models, it is necessary to
 solve equations (\ref{u2}) and (\ref{u5}) in terms of $\Theta$ so that
 specific values of $R$ and its derivatives are evaluated. In order to
 do so numerically, we need to find the interval where the root lies. It
 is not difficult to show (see paper I) that in the elliptic models
 \begin{equation}
     \frac{4}{F} ( t+\beta ) {\mid f^2-1 \mid }^{3/2} -1 
     \leq  2 \Theta  \leq \frac{4}{F} ( t+\beta ) 
      {\mid f^2-1 \mid }^{3/2} +1,
 \label{u6}
 \end{equation}
 and in the hyperbolic case
 \begin{equation}
      0 < \Theta \leq {\left[ \frac{3}{F} (t+ \beta ){(f^2-1)}^{3/2}
          \right] }^{1/3}.
  \label{u7}
 \end{equation}
 
 The local density in Tolman's spacetime is given by
 \begin{equation}
    8 \pi \rho = \frac{F'}{2 R' R^2}
  \label{t3}
 \end{equation}
  and if we adopt the radius coordinate $r$ as the parameter along the
  backward null cone ($r \ge 0$), we can then write the past radial null 
  geodesic of metric (\ref{t1}) as
  \begin{equation}
  \frac{dt}{dr} = - \frac{R'}{f}.
  \label{t4}
  \end{equation} 

The Friedmann metric is a special case of the metric
(\ref{t1}) above, obtained when $R = a(t) \ g(r)$, $f=g'$ and $\beta=$
constant. The usual Friedmann universe requires further that $g= \sin r$, 
$r$, $\sinh r$ and $F= b_1 \sin^3 r$, $\frac{8}{9} r^3$, $ b_2 \sinh^3 r$,
which are respectively the cases for $K=+1$, $0$, $-1$. The positive
constants $b_1$ and $b_2$ are scaling factors necessary to make the
density parameter $\Omega$ equal to any value different from one in the
open and closed models. It is also possible to
show that $F(r)$ gives the gravitational mass inside the comoving radius
$r$ and $f(r)$ gives the total energy of the system also within $r$ (see
\S 6.1 and paper II for further details).

It was shown in paper II how the Hubble constant relates to the
function $\beta(r)$ in a Einstein-de Sitter universe. It is important
here to extend that result to the $K= \pm 1$ Friedmann cases. By
definition, $H= \dot{R} / R = \dot{a} / a$ for constant $\beta(r)$.
Considering the especializations above, that permit us to get the Friedmann
metric from the Tolman solution, and assuming $t=0$ as our ``now'', that
is, $t=0$ being the time coordinate label for the present epoch, it is
straightforward to show that the present value $H_0$ for the Hubble
constant in the closed Friedmann model is given by
\begin{equation}
 H_0 = \frac{4 \sin 2 \Theta_0}{b_1 {(1- \cos 2 \Theta_0 )}^2},
 \label{add-1}
\end{equation}
where $\Theta_0$ is the solution of
\begin{equation}
 4 \beta_0 = b_1 (2 \Theta_0 - \sin 2 \Theta_0),
 \label{add-2}
\end{equation}
and $\beta(r)=\beta_0$ is a constant that gives the age of the universe.
In the open Friedmann case we will have then 
\begin{equation}
 H_0 = \frac{4 \sinh 2 \Theta_0}{b_2 {(\cosh 2 \Theta_0 - 1 )}^2},
 \label{add-3}
\end{equation}
and 
\begin{equation}
  4 \beta_0 = b_2 ( \sinh 2 \Theta_0 - 2 \Theta_0).
  \label{add-4}
\end{equation}

It will be necessary in further calculations to obtain the value of
the cosmological density parameter $\Omega$ in the Friedmann model at
the present constant time hypersurface. By definition
$\Omega = \rho / \rho_c$ and $\rho_c = 1/ (6 \pi {\beta_0}^2)$
at $t=0$. Therefore, in the closed Friedmann model we have that
\begin{equation}
  \Omega_0 = \frac{72 {\beta_0}^2}{ {(b_1)}^2 {(1- \cos 2
  \Theta_0)}^3},
  \label{add-5}
\end{equation}
where $\Theta_0$ is given by equation (\ref{add-2}). In the open
Friedmann model we have
\begin{equation}
  \Omega_0 = \frac{72 {\beta_0}^2}{ {(b_2)}^2 {(\cosh 2 \Theta_0 - 1)}^3},
  \label{add-6}
\end{equation}
with $\Theta_0$ being the solution of equation (\ref{add-4}).

It was shown in paper I that in a Tolman universe the redshift may
be written as
\begin{equation}
  1+z = {(1-I)}^{-1},
  \label{t5}
\end{equation}
where the function $I(r)$ is the solution of the differential equation
\begin{equation}
  \frac{dI}{dr} = \frac{\dot{R'}}{f} (1-I).
  \label{t6}
\end{equation}
The luminosity distance $d_l$ and the cumulative number count $N_c$
are given by
\begin{equation}
   d_l = R{(1+z)}^2,
   \label{t7}
\end{equation}
\begin{equation}
  N_c=\frac{1}{4 M_G} \int \frac{F'}{f} dr,
  \label{t8}
\end{equation}
where $M_G$ is the average galactic rest mass ($ \sim 10^{11}
M_{\odot}$). The volume $V$ of the sphere which contains the sources
and the volume density (average density) $\rho_v$ have the form
\begin{equation}
   V=\frac{4}{3} \pi {(d_l)}^3,
   \label{t9}
\end{equation}
\begin{equation}
   \rho_v = \frac{N_c M_G}{V}.
   \label{t10}
\end{equation} 
The proposed relativistic version of Pietronero's (1987) generalized
mass-length relation is given by 
\begin{equation}
   N_c = \sigma {(d_l)}^D
   \label{t11}
\end{equation}    
where $\sigma$ is a constant related to the lower cutoff of the
fractal system and $D$ is its fractal dimension. By substituting
equations (\ref{t9}) and (\ref{t11}) into equation (\ref{t10}) we
get the de Vaucouleurs' density power law
\begin{equation} 
   \rho_v = \frac{3 \sigma M_G}{4 \pi} {(d_l)}^{- \gamma}, \ \ \ \
   \gamma = 3-D.
   \label{t12}
\end{equation}  

\section{Numerical Analysis}
 \subsection{Initial Conditions}

The numerical task consists of finding solutions of
the geodesic equation (\ref{t4}) and the equation (\ref{t6}) that
enables the calculation of the redshift. Let us call the solution
of the former $t(r)$ and the solution of the latter $I(r)$. As we
are seeking results along the past light cone,
at each step of the independent variable $r_i$, we need to find $t_i$
first and then use it to find $I_i$ ($i=1,2,\ldots,n$).

 In paper I it was initially proposed to start the integration at $t(0)=0$
 and $I(0)=0$, but realized that in this approach there would be
 problems in the elliptic and hyperbolic models because we would get
 $\frac{0}{0}$ type indeterminations in the function $R(r,t)$ at $r=0$.
 In order to overcome this difficulty I then proposed a flat metric
 from $r=0$ till $r= \varepsilon$ ($\varepsilon$ small), beyond which
 the spacetime changed to a curved one. This hypothesis effectively
 replaces the initial conditions above by $t(\varepsilon)= -
 \varepsilon$, $I(\varepsilon)=0$. As we shall see, these new initial
 conditions produce numerical results which are perfectly
 acceptable~\footnote { \ This is especially true as in this
 problem we are seeking
 a fractal density profile, and, so, what is relevant to the problem
 are the intermediary results of the integration and not its final
 values.} and numerically stable, but in this approach the formulation
 and solution of the interface between flat and curved spacetimes is
 left incomplete 
 inasmuch as a full determination of this interface demands the solution
 of the junction conditions between the two regions. If this matching is
 not fully solved, discontinuities might arise across the joining
 surface, mainly in the function $R(r,t)$ which determines  the
 luminosity distance. As we are assuming $\varepsilon$ small (at least
 $10^{-3}$) these possible discontinuities do not necessarily represent a
 hazard to the numerical solutions, but we need to have a way to check
 them.

 The match between flat and Tolman metrics is intimately linked to
 the regularity conditions of the latter and, for the sake of clarity
 and completeness of this work, I shall discuss briefly the basic argument
 for Tolman regularity at $r=0$ (Bonnor 1974), with
 some additional remarks. Considering a displacement in the 2-surface
 $t=$ constant, $\phi=$ constant, equation (\ref{t1}) becomes
  \[
   -dS^2=\frac{{R'}^2}{f^2} \left( dr^2 + \frac{R^2 f^2}{R'^2} d \theta^2
                          \right).
  \]
 Near $r=0$ this 2-surface must be Euclidean, therefore, ${R'}^2 f^{-2}
 \rightarrow$ constant ($\not= 0$) and
   \[
     \frac{f^2 R^2}{{R'}^2} \sim r^2,
   \]
  that is,
 \begin{equation}
    \lim_{r \rightarrow 0} \frac{{R'}^2 r^2}{f^2 R^2} = 1,
    \label{t13}
 \end{equation}  
 which is Bonnor's (1974) equation (2.8). Now, if we suppose that  $f$
 is approximately constant for small $r$, equation ({\ref{t13}) can be
 integrated to get $R = r^f$. Therefore,
  \begin{equation}
    \lim_{r \rightarrow 0} \frac {R'}{f} = \lim_{r \rightarrow 0} r^{f-1}.
    \label{t14}
  \end{equation}
If $f < 1$ the limit above becomes infinity. If $f > 1$ the limit is
zero. Hence, $R' f^{-1}$ can only converge to a finite and non-zero
constant if $f=1$ as $r \rightarrow 0$, and $R \sim r$ also near $r=0$.
So the function $f(r)$ as an exponent in equation (\ref{t14}) and with 
the value $f=1$ appears to assume a role of {\it criticality} in regular
Tolman solutions.

To calculate the matching between flat and Tolman spacetimes, let
us start with two Tolman regions
\begin{eqnarray}
  {(d S_1)}^2 & = & dt^2 - {({R'}_1)}^2 {(f_1)}^{-2} dr^2 -
		     {(R_1)}^2 d \Omega^2; \ \ \ \ r < \varepsilon; 
                 \label{t15} \\
  {(d S_2)}^2 & = & dt^2 - {({R'}_2)}^2 {(f_2)}^{-2} dr^2 - 
		     {(R_2)}^2 d \Omega^2; \ \ \ \ r > \varepsilon. 
                \label{t16}             
\end{eqnarray}
The Darmois junction conditions at the comoving surface $r=\varepsilon$
were obtained by Bonnor \& Chamorro (1990) as 
\begin{equation}
  R_1(\varepsilon,t) = R_2(\varepsilon,t); \ \ \
  f_1(\varepsilon)   = f_2(\varepsilon); \ \ \
  F_1(\varepsilon)   = F_2(\varepsilon).
  \label{t17}
\end{equation}
If we take region 1 as flat with static boundary, we can write $R_1=r$,
$f_1=1$, and assuming $F_1=F_2=F(r)$, provided $F(0)=0$, the junction
conditions (\ref{t17}) are equivalent to 
\begin{equation} 
  R_2[\varepsilon,t(\varepsilon)] = \varepsilon, \ \ \
  f_2(\varepsilon) = 1.
  \label{t18}
\end{equation}
This result means that on the curved side of the joining surface the solution
must be parabolic. Considering the parabolic solution (\ref{u3})
the matching conditions ({\ref{t18}) are equivalent to solving the
following equation for $\varepsilon$:
\begin{equation}
  9 F(\varepsilon) { \left[ \beta(\varepsilon) - \varepsilon \right] }^2
  - 8 \varepsilon^3 = 0, \ \ \ \ F(\varepsilon) \ge 0,
  \label{t19}
\end{equation}
where the geodesic for $r \le \varepsilon$ is the flat one $t=-r$.
Therefore, we can start the numerical integration with the initial
conditions
\begin{equation}
  t(\varepsilon) = - \varepsilon, \ \ \ I(\varepsilon) = 0,
  \label{t20}
\end{equation}
while  
\begin{equation} 
  R(\varepsilon) = \varepsilon, \ \ R'(\varepsilon) = 1, \ \
  \dot{R}(\varepsilon) = 0.
  \label{t21}
\end{equation}
It was the explicit values of the function $R[r,t(r)]$ on the 
surface $r=\varepsilon$ that were missing in the previous discussion made
in paper I.
Notice that the junction conditions do not require $R'$ to be continuous
across the joining surface.

It is still possible to simplify the problem by noticing that
$\varepsilon = 0$ is actually a solution of equation (\ref{t19}). Hence,
instead of initial values (\ref{t20}) and (\ref{t21}) we have
\begin{equation}
  t(0) = 0, \ \ \ \ I(0)=0,
  \label{t22}
\end{equation}
and
\begin{equation}
  R(0) = 0, \ \ R'(0) = 1, \ \ \dot{R}(0) = 0.
  \label{t23}
\end{equation}
In this way the previous difficulty of indetermination of the function
$R$ at $r=0$ is avoided as this method allows us to obtain explicit
values of $R$ at the origin. This is obviously a consequence of the
complete formulation and solution of the interface between the two spacetimes.

The initial conditions (\ref{t22}) and (\ref{t23}) may be obtained
alternatively without mention of the flat-curved junction conditions.
Around $r=0$ the regularity conditions $f(0)=1$, $F(0)=0$ reduce the
field equation (\ref{t2}) to simply $\dot{R}=0$. As we know that
$R \sim r$ when $r \rightarrow 0$, we obtain at once the two other values
in equations (\ref{t23}).

In conclusion, we have in practice three sets of initial
conditions, where each one, in theory, enables the numerical integration
of equations (\ref{t4}) and (\ref{t6}):
\begin{namelist}{iiixx}
 \item[{\it (i)}] the initial values (\ref{t20}) and (\ref{t21}) with
                  $\varepsilon$ being found by solving numerically
		  equation (\ref{t19});
 \item[{\it (ii)}] the initial conditions (\ref{t22}) and (\ref{t23});
 \item[{\it (iii)}] assuming initial values (\ref{t20}) and taking
               %  $\varepsilon$ small ($\leapp 10^{-3}$), but calculating
		  $\varepsilon$ small ($\stackrel{<}{\sim} 10^{-3}$), but calculating
		  $R(\varepsilon)$, $R'(\varepsilon)$, and
		  $\dot{R}(\varepsilon)$ by putting $\varepsilon$
		  directly into equations (\ref{u1}),
		  (\ref{u3}) and (\ref{u4}) and their derivatives.
\end{namelist}
This third procedure implies neglecting the results
(\ref{t21}) of the matching and  that will almost surely give rise to
discontinuities in $R(r,t)$. Nevertheless, provided $\varepsilon$ is
sufficiently small (but not too small otherwise $R$ blows up) and
equations (\ref{t4}) and (\ref{t6}) are numerically stable, we would
usually get in the end the equivalent results as if we had used initial values
(\ref{t22}) and (\ref{t23}) with a small perturbation $\varepsilon$. As we
shall see next, this last remark will prove to be most valuable in
acquiring confidence in the numerical results.

 \subsection{Numerical Algorithm}

Once the initial value problem for the numerical integration of
the first-order ordinary differential equations (\ref{t4}) and
(\ref{t6}) is determined, the next step is to choose the numerical
integrator that propagates the solution over the interval. In this
respect it is essential to bear in mind that what is being sought is a
qualitative fractal behaviour for the solutions, that is, solutions
which graphically follow the de Vaucouleurs density power law
(\ref{t12}). Therefore, high accuracy is not important in this problem.
In addition, although the problem is mathematically tricky, the numerical
tasks are relatively simple and, so, computational efficiency is of no
great concern. In view of this the explicit fourth-order Runge-Kutta
method with adaptive stepsize control was the integrator chosen
(Press et al. 1986).

As discussed in paper I, the aim of modelling a fractal distribution
by Tolman's solution is to make use of the freedom of the arbitrary
functions in order to ascertain particular functions $f(r)$, $F(r)$,
$\beta(r)$ such that the volume density takes the de Vaucouleurs density
power law (\ref{t12}). The way to check whether a fractal distribution
was modelled is by taking the logarithm of equation (\ref{t12})
\begin{equation}
 \log \rho_v = a_1 + a_2 \log d_l
 \label{u8}
\end{equation}
and finding the constants $a_1$ and $a_2$ by linear fitting over the
points obtained through numerical integration. The fitting is considered
a success if it is linear with a negative slope. That can be concluded
by visual inspection of the graph and by an acceptable goodness of fit.
Inasmuch as what we are seeking is a qualitative fractal behaviour, very
often a visual inspection will be enough.

I shall  describe below in general terms the algorithm used to
apply the model and obtain the numerical results desired.
\begin{namelist}{10xx}
 \item[{\rm 1)}] START by choosing special forms for the functions
		 $f(r)$, $F(r)$, $\beta(r)$, such that $f(0)=1$ and $F(0)=0$;
 \item[{\rm 2)}] discretize the radial coordinate $r_i$ $(i=1,2,
		 \ldots,n-1,n; \ \varepsilon \le r \le \Sigma_0; \
		 r_1=\varepsilon, \ r_n=\Sigma_0$; total of $n-1$
		 steps)\footnote{ \ In practice the discretization $r_i$ will
		 be performed automatically by
		 the adaptive stepsize control routine which keeps the
		 accuracy of the numerical results within a desired
		 predetermined value (Press et al. 1986).};
 \item[{\rm 3)}] make $i=1$ and choose a set of initial conditions to get
		 $t_1$ and $I_1$;
 \item[{\rm 4)}] evaluate $f_i$, $F_i$, $\beta_i$, ${f'}_i$, ${F'}_i$,
		 ${\beta'}_i$;
 \item[{\rm 5)}] if ${f_i}^2 > 1$ or ${f_i}^2 < 1$ then evaluate $\Theta_i$
		 and consider $t_i$ to calculate $R_i$, $R'_i$,
		 $\dot{R'}_i$; else evaluate directly $R_i$, $R'_i$,
		 $\dot{R'}_i$ considering $t_i$;
 \item[{\rm 6)}] evaluate $z_i$, $d_{l_i}$, $N_{c_i}$, $V_i$, $\rho_{v_i}$;
 \item[{\rm 7)}] by using the numerical integrator chosen, advance solution
                 $t_{i+1}$ then use value $t_i$ to advance solution $I_{i+1}$;
 \item[{\rm 8)}] make $j=i+1$ then $i=j$; if $i=n$ then continue below;
		 else return to stage 4;
 \item[{\rm 9)}] make final evaluations $f_n$, $F_n$, $R_n$, $R'_n$,
                 $z_n$, etc;
 \item[{\rm 10)}] with the collection of points $\rho_{v_k}$ and $d_{l_k}$
                  ($k=2,3, \ldots,n$) perform a $n-1$ points linear
		  fitting according to equation (\ref{u8});
 \item[{\rm 11)}] if fitting is unsuccessful then discard used functions $f(r)$,
		  $F(r)$, $\beta(r)$ and go to stage 1; else STOP.
\end{namelist}

The computer routine which performs this algorithm was written in 
standard Fortran 77 in double precision and the interested reader can
find it in Ribeiro (1992c). Each run of the routine, either successful or
not, constituted a different numerical simulation. The program uses many
subroutines of Press et al (1986) with some minor modifications.
Round-off and truncation errors are monitored at each step by
calculating the deviation from zero of the field equation
(\ref{t2}) and its first partial derivative in $r$. Stability and
accumulation of errors are monitored by taking the final results as
initial conditions to
reverse the whole integration and compare the results with the initial
values. Error propagation formulas are also used to see whether any of
the derived observational quantities acquires unacceptably high errors.
The units too are chosen bearing in mind the need to minimize errors
and avoid large numbers. Hence, distance is given in Gpc ($10^9$ pc) and in
order to have $c=G=1$, time is expressed in 3.26 Gyr units
(1~yr~$=~3.16~\times~10^7$~s) and mass is expressed in units of
$2.09 \times 10^{22} M_{\odot}$.

The first numerical simulations used the set {\it (iii)} of initial
conditions described above with $\varepsilon = 10^{-3}$. The fractional
accuracy for the local truncation errors in the variable-step routine
was set at $10^{-4}$, and this was the value in all simulations, that
is, the results throughout this paper were all obtained with this
accuracy. This set {\it (iii)} of initial conditions applied to
equations (\ref{t4}) and (\ref{t6}) led to numerically stable results,
with the routine taking a relatively small number of steps to go through
the interval of integration. Around 10 to 20 successful steps  were
usually enough to cover that interval, with 1 or 2 failed steps (see
Press et al. 1986, p. 555, for details of the workings of an adaptive
stepsize control routine). The reverse integration, that is, taking the
final results to integrate backwards in order to compare with the
initial conditions, produced a few more failed steps, but the comparison
with the initial conditions showed results well within the desired
accuracy. These details are important to highlight the smoothness and
stability of the solutions, although this approach actually gave rise to
discontinuities in $R$ across the surface $r = \varepsilon$, as
expected (more details about the way the routine works and its output are
available in Ribeiro 1992c).

As far as numerical stability is concerned, the situation changed
dramatically when the set {\it (ii)} of initial conditions was used,
that is, taking $\varepsilon = 0$ and assuming initial values (\ref{t22})
and (\ref{t23}). The solutions became unstable about the origin, with
the program being forced to decrease the stepsize so much that it needed
to take on the order of hundreds of steps to cover a tiny interval very
close to the origin. Once they emerged from this unstable region, the
solutions started
behaving like the previous approach, covering the remaining region very
quickly. The reverse integration behaved likewise, with hundreds
of failed steps. Despite this instability the solutions were
still very smooth and the comparison of the initial conditions with the
results of the backward integration were still well within the desired
accuracy. Thus it would appear that with this approach the numerical
problem is {\it stiff} across the joining surface.

Integrating a stiff problem with an explicit Runge-Kutta adaptive stepsize
control routine greatly reduces the efficiency of the code as the initial
stepsize chosen puts the method at or near numerical instability, which
causes a large truncation error estimate. That leads the routine to
reduce substantially the stepsize to keep the local truncation error
within the predetermined accuracy. Then the method is able to successfully
integrate the problem, but uses a far smaller stepsize and a far greater
number of steps than seems justified if we consider the smoothness of
the solution (Lambert 1991).

Despite the loss of efficiency, our simulations were usually
sufficiently straightforward and computation time did not pose a barrier
in most of them. In addition, double precision usually provided enough
manoeuvring space for the decrease of stepsize. The situation, however,
was far worse when the simulations were carried out using the set
{\it (i)} of initial conditions, that is, with $\varepsilon \not= 0$.
The stiffness was so severe that after a few iterations the step length
became so small that the difference $r_{i+1} - r_i$ went
beyond the machine's precision and the routine achieved an apparently zero
stepsize. Obviously such a case cannot be integrated unless an implicit method
is used or, perhaps, if a quadruple precision routine is implemented in
our explicit Runge-Kutta method. In any case, as the set {\it (ii)} of
initial conditions provides a simpler approach and allows the
stiffness to be handled by the program in most cases of interest, these
were the conditions used to obtain all results shown in this paper,
unless stated otherwise.

It is important to mention that when comparing the results obtained
by using the set {\it (ii)} with the ones produced by the set
{\it (iii)}, the final results of both integrations usually differ by about
$\left| \varepsilon \right|$, which substantiates the interpretation
at the end of \S 3.1 that the results obtained with initial
conditions {\it (iii)} are usually equivalent to the ones obtained with
conditions {\it (ii)}, but perturbed by about $\varepsilon$ (see Ribeiro
1992c for an example of such a situation).

Stiffness is associated with the existence of transient terms in
the solution: they contribute very little to the solution, but the usual
methods require that they be approximated accurately in order to maintain
stability (Gear 1971; Ortega \& Poole 1981). In this case it appears that
the imposition of the regularity conditions upon the Tolman solution, or
the similar process of joining it to a flat region, give rise to these
transient terms. Nonetheless, the unanswered question is whether the
Tolman stiffness is a mere numerical phenomenon in the solutions, with
no major implications, or if it indicates a deeper physical reason in
the Tolman models. It is known that some chaotic dynamical systems
present stiff behaviour, but the interaction between stiffness and chaos
is still unclear (Cartwright \& Piro 1992). More work is necessary in
order to clarify this point.

\section{The Spatially Homogeneous Special Cases}

In this section I shall present the results of the integration of
the spatially homogeneous Friedmann models, obtained as special cases of
the Tolman metric. The aim of this section is twofold: first to extend
the analysis of the Einstein-de Sitter model presented in paper II to
the $K = \pm 1$ Friedmann cases, and second to provide numerical results
of the Einstein-de Sitter case that can be compared with the exact
solution already presented in paper II.

As discussed in \S 2 the Einstein-de Sitter model is obtained as
a special case of the Tolman solution when $f=1$, $F= \frac{8}{9} r^3$,
$\beta = \beta_0$, where $\beta_0$ is a constant, and this assumption means 
that along the past light cone the volume density and luminosity 
distance are given by (see paper II)
\[
  \rho_v = \frac{{ ( 3 {\beta_0}^{1/3} - r ) }^6}
                {54 \pi { ( 3 \beta_0 ) }^4},
\]
\[
  d_l = \frac{9 r {\beta_0}^{4/3}}{{ ( 3 {\beta_0}^{1/3} - r ) }^2}.
\]
Throughout this section
I shall assume the value $75 \ \mbox{km} \ {\mbox{s}}^{-1} \
{\mbox{Mpc}}^{-1}$ for the Hubble constant, and this means $\beta_0=2.7$
for the Einstein-de Sitter model in our adopted units. Figure 1 shows
the $\log \rho_v$ versus $\log d_l$ graph obtained when using the 
analytical expressions above in the parameter range $0.001 \le r \le 1.5$,
and figure 2 shows the same results obtained by integrating numerically
the Einstein-de Sitter model in the same parameter range. By
comparing the two graphs one can see that the numerical integration does
reproduce the analytical results. Figures 3 and 4 show
respectively the null geodesic $t(r)$ and the function $I(r)$ obtained
numerically for this case, and there the smoothness and almost linearity
of the solutions within the range of integration are clearly visible.
The analytical expressions for these two functions, obtained in
paper II, may be written as
\[
  t= { \left( {\beta_0}^{1/3} - \frac{r}{3} \right) }^3 - \beta_0,
\]
\[
 I= 1- { \left( 1 - \frac{r}{3 {\beta_0}^{1/3}} \right) }^2,
\]
and by means of a few direct numerical substitutions one can see that the
numerical results shown in figures 3 and 4 do reproduce the analytical 
functions above, as expected.

The Einstein-de Sitter model also
exhibits stiffness, but this is not visible in figure 2 because only
steps bigger than $0.0025$ were stored in the integration shown in that
figure. Figure 5, however, shows
the same integration of figure 2, but with all points used by the code
being printed. The stiffness is then clearly visible as the program takes
an excessive number of steps around $r=0$ despite the smoothness and
linearity of the solutions $t(r)$ and $I(r)$ in the range of the
Einstein-de Sitter model shown in figures 3 and 4.

The open Friedmann model with $\Omega_0 \cong 0.2$ is obtained by 
taking $f=\cosh r$, $F= \sinh^3 r$ and $\beta_0= 3.6$. Figure 6 shows
the numerical results of
$\log \rho_v$ against $\log d_l$ for this model, also integrated in the
range $0 \le r \le 1.5$. By comparing figure 6 with figure 2 we can see
that in the open Friedmann model $\rho_v$ departs from a constant value at
about $\log d_l = -0.6 \ (d_l \approx 250 \ \mbox{Mpc})$, while in the
Einstein-de Sitter case that happens at about $\log d_l = -1 \ (d_l
\approx 100 \ \mbox{Mpc})$. In other words, the departure from a homogeneous
distribution, where $\rho_v$ is constant, is deeper in the open
Friedmann model. Stiffness is also present in this case, but in a more
severe form as the code took a far greater number of steps for the
integration to emerge from around the origin than in the equivalent 
Einstein-de Sitter integrations.

Finally, figure 7 shows the graph for the recollapsing closed 
Friedmann model with $\Omega_0 \cong 4$, which is obtained as a special
case from the Tolman
metric when $f=\cos r$, $F=\sin^3 r$, $\beta_0= 0.7$. The numerical
integrations were also carried out in the same parameter range as the
other two cases, and figure 7 shows the results of $\rho_v$ vs. $d_l$.
In this case the volume density starts to
change from a constant value at around $\log d_l = -1.5 \ (d_l \approx
30 \ \mbox{Mpc})$. Therefore, in the recollapsing model the integration
along the backward null cone reaches different and earlier spatial sections at
closer distances than in the other two spatially homogeneous models.
Notice that the values of the luminosity distance where the departure
from homogeneity begins are dependent on the Hubble constant used in
these integrations, and, therefore, the numbers given above should be
considered as rough approximations.

The recollapsing Friedmann model could not be integrated with the
set {\it (ii)} of initial conditions because the stiffness led to very
small first steps, where $F \approx 0$ and $f \approx 1$ when
$r\rightarrow 0$. Hence, the two limits of the interval given by equation
(\ref{u6}) quickly became extremely big numbers and finding $\Theta$ by
means of equation (\ref{u6}) broke down due to the machine's precision
limitations. Besides, the numerical instability often led to either shell
crossing behaviour at very small $r$ or to unacceptably high error
accumulation, and in consequence this numerical approach could not be
relied upon. As a result the recollapsing integrations were carried out
using the set {\it(iii)} of initial conditions with $\varepsilon = 10^{-3}$.

The results for the $K= \pm 1$ Friedmann cases show that they
also do not appear to remain homogeneous along the backward
null cone. In addition, considering the form of the graphs of these two
cases we can conclude that similarly to the Einstein-de Sitter
case studied in paper II, the recollapsing and open Friedmann models do 
not have single fractal features along the past light cone as they do not
follow the de Vaucouleurs' density power law (\ref{t12}). Therefore,
considering the IRAS data, all three spatially homogeneous Friedmann 
models appear to have difficulties in modelling those observations as
they predict homogeneity where it is not observed, and for deeper
regions all models start to deviate from it.

The apparent inhomogeneity of the Friedmann models is a consequence of the fact
that in the present approach densities are expressed along the past light
cone, which is where observations are actually made. This means that the
volume density $\rho_v$ goes through hypersurfaces of constant $t$ where
each one has different values for the proper density. Therefore, although
the Friedmann models are {\it spatially} homogeneous, the volume density
will only be homogeneous, that is, have a constant value, in local regions.
This will only happen while it is being calculated nearby our present
constant time hypersurface, and once we depart from its neighbourhood
the volume density starts to change, with the model becoming increasingly
apparently inhomogeneous the
further the calculations go into the past. The change of the volume density 
along the geodesic is due to it being a cumulative density,
averaging at bigger and bigger volumes in a manner that adds more and
more different local densities of each spatial section of the models.
Therefore, under this definition the ``homogeneity'' of the Friedmann
models does not survive. It is important to realize that this interpretation
of the ``homogeneous'' Friedmann model is closely related to the fractal
approach to cosmology adopted in this series of papers, inasmuch as the
definition for the volume density used here is a direct consequence of
the assumption that the large-scale structure of the universe can be
described as being a self-similar fractal system.

As final remarks, when one approaches the singularity the model is usually
said to become denser because the local density $\rho$ tends to infinity. 
However, this is {\it not} the case for the volume density $\rho_v$,
which vanishes at the big bang. This behaviour definitively happens in
the Einstein-de Sitter model (see paper II for the asymptotic analysis
of this model) and due to the similar form of the graphs of
$\log~\rho_v$~vs.~$\log~d_l$ in the open and closed Friedmann models (see
figures 6 and 7) as compared to the Einstein-de Sitter case (figures 1 and 2),
we can then assume that this vanishing volume density is a feature of all
Friedmann models when they approach the singularity. Such a behaviour for
$\rho_v$ is, once again, explained by its adopted definition
[eq. (\ref{t10})]: at the big bang singularity hypersurface the observed
volume (as defined through the luminosity distance) is infinite, but the
total mass within it
is finite. It is interesting to note that Wertz (1970) had already
postulated a vanishing global density as a requirement for the so-called
Pure Hierarchical Models, and similarly Pietronero (1987) conjectured
the same sort of result for a fractal distribution. Therefore, it does 
seem valid to say that at least some aspects of a hierarchical (fractal)
cosmology can already be found {\it within} the standard Friedmannian
cosmology, provided one starts with the appropriate definitions.

Empirically, the effect of zero global density in these models as they
approach the big bang can be seen as the consequence of the fact that in
this work the galaxies one sees empirically on the past light cone are
plotted at their luminosity distances, and the luminosity distances go
to infinity as $z$ goes to infinity while the number of galaxies remains
finite. Observers do often plot their data in ways like that, and an
example is Geller \& Huchra's (1989) slices where galaxies are plotted
at a distance proportional to their redshift $z$. If they had extended
their plot to $z$ equals infinity only a finite number of galaxies would
be inside the horizon and one would see the galaxies thin out to zero.
It is obvious, however, that if a different definition of distance had
been used (say comoving distance), this effect of zero global density at
the big bang singularity would no longer be necessarily found,
nevertheless the redshift or the luminosity distance are observable
quantities, independent of cosmology, and observers do use them.
Therefore, we can then consider the de Vaucouleurs' density power law as
being essentially an empirically observed relation when observables are
plotted, and from this perspective this work can be viewed as an attempt
to see whether there are any inhomogeneous big bang cosmologies that
would allow this empirical relation to be extended to larger scales.

\section{Tolman Fractal Solutions}

Before discussing the solutions themselves, let us first make some
remarks about the three arbitrary functions which may help to simplify
the simulations. The function $F(r)$ cannot be a constant, otherwise due
to the regularity condition $F(0)=0$ this constant would have to be zero
and there would not be any dust in the model. In addition, due to
equation (\ref{t3}) we have to restrict ourselves to $F' \ge 0$,
otherwise we would get negative mass in the model. Hence $F(r)$ must be
a positive and monotonically increasing function for $r \ge 0$. Notice,
however, that there could be ranges for $r$ where $F(r)$ is a nonzero
constant and in those ranges $F'=0$, implying the existence
of vacuum regions in the model. The only restriction for the function
$\beta(r)$ is that its values must be such that the model is kept within
its physical region, that is, such that $t + \beta > 0$. The aim of the
simulations is to find a qualitative fractal behaviour which could be
expressed in terms of a relatively simple metric. Therefore, it is
desirable to obtain fractal behaviour in terms of simple functions.
So, wherever possible, I shall restrict the choice of $f(r)$ to the
forms it takes in the spatially homogeneous cases, that is, specializing
only to $f = \cos r, \ 1, \ \cosh r$.

In deciding which solutions are acceptable as fractals, we have the
mathematical criterion of verifying whether they follow the equation
(\ref{t12}). However, some basic observational constraints must also be obeyed
in order to have physically realistic models. The first and most
obvious is the observed linearity of the redshift-distance relation
for $z \stackrel{<}{\sim} 1$. The second is the value of the Hubble constant
itself which, considering the present uncertainty in its measurements, is
accepted to be in the range $40 \ \mbox{km/s/Mpc} < H_0 < 100
\ \mbox{km/s/Mpc}$ (see White 1990 for a recent account on the subject).
Finally we have a constraint in the fractal dimension of the
distribution. Considering Pietronero's (1987) analysis based on the
measurements of the 2-point correlation function, we would have
$D \approx 1.4$. However, as other authors claim different and higher
values for $D$ (Deng, Wen \& Liu 1988; Calzetti \& Giavalisco 1991), I
shall loosely assume here that the wider range $1 < D < 2$ is reasonable
enough for the purposes of this work.

 \subsection{Parabolic Solutions}

The parabolic class of Tolman models is the natural one to start
the search for fractal solutions, inasmuch as one of the three functions
is already specified, that is, $f=1$ in this class. Nevertheless, and
despite the simplifying remarks above, it was not at all an easy task to find
those results. It took over 250 different simulations to find suitable
solutions of parabolic type, and one of the major problems was to find
the range of the constants in the functions where the overall solution
does have a power law behaviour within the interval of integration.

Despite those difficulties, fractal parabolic solutions along the
past light cone do exist, although I could only find them by
assuming a non-simultaneous big bang, that is, I could not find
fractal results with constant $\beta(r)$. Below are the particular
forms of the remaining arbitrary functions that led to fractal
behaviour in parabolic models:
\begin{equation}
 \left\{ \begin{array}{ll}
	 F = \alpha r^p, \\
	 \beta = \beta_0 + \eta_0 \ r^q,
	 \end{array}
 \right.
 \label{t24}
\end{equation}
and
\begin{equation}
 \left\{ \begin{array}{ll}
	  F = \alpha r^p, \\
          \beta = \ln \left( {e}^{\beta_0} + \eta_1 r \right),
         \end{array}
 \right.  
 \label{t25}
\end{equation}
where $\alpha$, $p$, $q$, $\beta_0$, $\eta_0$, $\eta_1$ are positive
constants. \footnote{ \ Actually, by a coordinate transformation $F$ can
be made equal to $\frac{8}{9}r^3$, so the only genuine arbitrary function
is $\beta$.} The simulations showed that in both cases above $\alpha$ must
be around $10^{-4}$ to $10^{-5}$ and $p$ and $\beta_0$ can vary from
around 0.5 to 4. For functions (\ref{t24}) $q$ lies around 0.65 and
$\eta_0$ around 50. For solutions (\ref{t25}) $\eta_1$ varies from about 1000
to 1300. It is difficult to attribute specific roles for each constant as
usually a change in one of them changes the whole behaviour of the
solution, but we can very roughly say that  $\alpha$, $q$, $\eta_0$ and
$\eta_1$ are more related to the linearity of equation (\ref{u8}) and
that $p$ and $\beta_0$ are more related to its slope. In other words,
changing $p$ and $\beta_0$ affects more sharply the fractal dimension of
the solutions, but without changing too much their linearity. In
addition, if in equations (\ref{t25}) we use instead $\beta= \ln \left[
{e}^{\beta_0} + \eta_1 \left(r+r^2 \right) \right]$, the linearity of
the solutions is slightly improved.

Figure 8 presents the results of an integration obtained with 
functions (\ref{t24}). The fractal dimension is $D=1.4$ and the
integration was carried out in the interval $0 \le r \le 2$. In figure 9
one can see the numerical results for the functions (\ref{t25})
integrated in the range $0 \le r \le 5$, with the resulting fractal
dimension $D=1.7$. Both figures also show the straight lines fitted
according to equation (\ref{u8}) and the qualitative power law nature
of the solutions in the integrated interval is visually obvious.
We can also see from the figures that the two solutions actually are not
much different from one another.

These solutions, however, fail to meet the other observational
criteria outlined above. Figure 10 shows the $d_l$ vs. $z$ plot of
results of the integrations of functions (\ref{t24}) and the
non-linearity of the relation is clearly visible. In addition, the associated 
Hubble constant is far too low and these reasons seem to rule out as
observationally unrealistic a model based on the functions (\ref{t24}).
Figure 11 presents the same results for functions (\ref{t25}) and
one can see that the redshift-distance relation is mostly linear,
especially for the region $z \ge 0.011$. The Hubble constant
associated with the linear part of the diagram is around 13, far below 
the lower limit for $H_0$ given above, and considering this point
it seems that the functions (\ref{t25}) also do not seem to produce
a realistic fractal model.

Two remarks must be made about the fractal parabolic solutions above. In
the first place, from a mathematical point of view the linearity of the
Hubble law~\footnote{ \ Here by Hubble law I mean the redshift-distance
law. See Harrison (1993) for a discussion about the distinction between
the Hubble law and the velocity-distance law.} must be approximately
valid through the origin, obeying the equation $d_l=c z/H_0$. This means
the redshift distance relation in figure 11 would not be acceptable even
though a section of the curve looks linear. However, from a physical
viewpoint one might argue that the linearity of the Hubble law is not
really needed very near the origin where it would anyway be messed up by
local peculiar velocities. In the second place, although it might be
argued that a different and more observationally acceptable value for
the Hubble constant could be obtained by changing the constants until the
desired value for $H_0$ is produced, it must be stressed that the fractal
behaviour of the solutions is sensitive (sometimes very sensitive) to
the choice of these same constants. In other words, the fractal
dimension itself and the limits over which the de Vaucouleurs density
power law applies are sensitive to the choice of these constants. 
Therefore, changing them in order to get a desired value for the Hubble
constant will only be succesful at the very likely cost of destroying
the fractal feature of the solutions over the desired observational
range.  Due to this reason such a procedure was ruled out in all 
simulations. \footnote{ \ This sensitivity of the fractal solutions to
small changes in the constants seems to indicate that the fractal
solutions presented in this paper appear to be structurally fragile
(Coley \& Tavakol 1992).}

As final remarks, although one cannot say that the functions
(\ref{t24}) and (\ref{t25}) are the only ones to produce fractal
parabolic solutions as there may be others which lead to fractal
behaviour but were not considered here, the experience with all
simulations makes it hard
to see how other solutions of this kind could be produced in terms of
simple functions. This point added to the unrealistic distance-redshift
relations of the models leads us to conclude that parabolic type Tolman
models do not appear to be able to produce physically realistic fractal
models in the backward null cone.

 \subsection{Elliptic Solutions}

The experience with parabolic models was valuable guidance for
further studies of Tolman fractal solutions. Elliptic models with
power law behaviour in the density were found directly from their
parabolic counterparts:
\begin{equation}
 \left\{ \begin{array}{ll}
	 f = \cos r, \\
	 F = \alpha r^p, \\
	 \beta = \beta_0 + \eta_0 \ r^q,
	 \end{array}
 \right.
 \label{t26}
\end{equation}
and
\begin{equation}
 \left\{ \begin{array}{ll}
	 f = \cos r, \\
	  F = \alpha r^p, \\
          \beta = \ln \left( {e}^{\beta_0} + \eta_1 r \right),
         \end{array}
 \right.  
 \label{t27}
\end{equation}
where $\alpha$, $p$, $q$, $\beta_0$, $\eta_0$, $\eta_1$ are positive
constants which must lie approximately in the same range as in
the parabolic cases in order to produce fractal solutions, \footnote{ \ Since
one can make a coordinate transformation in $r$, only two of the functions
$f$, $F$, $\beta$ are really arbitrary.} apart from $\alpha$ which must be
around the value 10. We have to remember that in the elliptic class models
the mass is higher than in the other two classes, which means that
the dust does not escape from its own gravitational field and this leads
to the eventual halt in the expansion when every part of the model starts
to contract. Therefore, it is reasonable to expect a higher value for
$\alpha$ in these cases.

The numerical integrations of all elliptic models were carried out
using the set {\it (iii)} of initial conditions as described in \S 3.1.
This was done for the same reason as in the recollapsing Friedmann
model: the set {\it (ii)} of initial conditions and equation (\ref{u6})
broke down due to severe numerical instability at very small $r$, which
produced unreliable results, if any.

Figure 12 shows the $d_l$ vs. $\rho_v$ results for functions
(\ref{t26}), with a resulting fractal dimension of $D=1.7$, and figure
13 shows the same results for functions (\ref{t27}) where $D=2$ was the
dimension found. Figure 14 shows the distance-redshift relation for the
model with equations (\ref{t26})  and we can see that the plot is not
linear. Finally, figure 15 shows the $d_l$ vs. $\rho_v$ plot for the
model of equations (\ref{t27}), and although the relation is more or
less linear it leads to the low value of roughly 36 for $H_0$.
Considering all those results together it also seems that equations
(\ref{t26}) and (\ref{t27}) produce models which are not compatible with
observations. Models with constant $\beta$ have less linear $\log
\rho_v$ vs. $\log d_l$ plots than the two previous ones and lead to too
high values for $H_0$ and $D$, making them also incompatible with
observations.

 \subsection{Hyperbolic Solutions}

The best fractal results in terms of power law behaviour for
$\rho_v$ vs. $d_l$ and agreement with observations were obtained by
numerical simulations of hyperbolic type solutions. The
specializations for $F(r)$ and $\beta(r)$ studied so far also lead
to power law behaviour in the density once the function $f(r)$ is
modified accordingly. Thus, we have the following sets of
functions: \footnote{ \ As in the elliptic case, here again only two
functions are really arbitrary.}

\begin{equation}
 \left\{ \begin{array}{ll}
	 f = \cosh r, \\
	 F = \alpha r^p, \\
	 \beta = \beta_0 + \eta_0 \ r^q,
	 \end{array}
 \right.
 \label{t28}
\end{equation}
and
\begin{equation}
 \left\{ \begin{array}{ll}
	 f = \cosh r, \\
	 F = \alpha r^p, \\
         \beta = \ln \left( {e}^{\beta_0} + \eta_1 r \right),
         \end{array}
 \right.  
 \label{t29}
\end{equation}
where, as before, $\alpha$, $p$, $q$, $\beta_0$, $\eta_0$, $\eta_1$
are positive constants. In this case the constants will assume
approximately the same
numerical ranges as in the parabolic case in order to produce fractal
solutions, apart from $p$ which for values smaller than 1.4 and greater
than 2.5 makes the problem too stiff to be integrated with initial
conditions {\it (ii)}. For those values the use of the initial
conditions {\it (iii)} will produce the desired results.

Figure 16 shows the $\rho_v$ vs. $d_l$ numerical results of the
integration of the model using equations (\ref{t28}) and the fitted
straight line, which leads to a fractal dimension $D=1.4$, while figure
17 shows the redshift-distance diagram. Although the latter relation 
is approximately linear, it leads to an $H_0 \approx 23$ km/s/Mpc, a
value too low to make equations (\ref{t28}) a model compatible with
observations.

The power law behaviour, leading to fractal dimension $D=1.3$, of
the results of the numerical integrations of the model formed by
functions (\ref{t29}) can be seen in figure 18. Figure 19 shows the
redshift-distance diagram of the same model and we can see 
that the plot is linearly well approximated. More important, a simple
calculation shows that the slope of the approximated line formed by
the points in figure 19 gives $H_0 \cong 61 $ km/s/Mpc. Therefore, the
model formed by functions (\ref{t29}) is the first so far to reasonably
agree with all basic requirements outlined above in order to make the model
compatible with current observations: linearity of the $d_l$ vs. $z$
diagram, Hubble constant within the currently accepted uncertainty, power
law behaviour for the $\rho_v$ vs. $d_l$ results and fractal dimension
around the values which are in agreement with present calculations
of the 2-point correlation function. Note incidentally, that the
integrations were stopped at $d_l \cong 350$ Mpc, which corresponds
to $z \cong 0.07$, as this is the redshift of the deepest self-similar
structures identified in the IRAS survey (Saunders et al. 1991). 

Even better results were obtained with the mathematically and
physically simpler model below:
\begin{equation}
 \left\{ \begin{array}{ll}
           f = \cosh r, \\
           F = \alpha r^p, \\
           \beta = \beta_0.
         \end{array}
 \right.
 \label{t30}
\end{equation}
This is obviously a simultaneous big bang model, like its Friedmann
counterpart, and is physically and mathematically the closest model to
the open Friedmann one, differing only in the function $F(r)$ which
gives the distribution of dust. Figure 20 shows the power law behaviour
of $\rho_v$ vs. $d_l$ of the model formed by functions (\ref{t30}), with a
fractal dimension of $D=1.4$. Figure 21 is the redshift-distance diagram
of the same model where we can see the good linear approximation given
by functions (\ref{t30}). The slope of the points gives $H_0 \cong 80$
km/s/Mpc, and it is interesting to note that recent measurements made
by two different methods suggest a Hubble constant very close to this value
(Peacock 1991). Actually, for $\beta_0 = 3.6$, the value used to get the
results shown in figures 20 and 21, we would have an age of the universe
of about 12 Gyr, which is a lower limit if we consider the age of
globular clusters (Peacock 1991). Therefore, also in this point of the
age of the universe the model (\ref{t30}) agrees reasonably well with
observations. The integrations with functions
(\ref{t30}) were also stopped at $z \cong 0.07$, which in this case
corresponds to the luminosity distance $d_l \cong 270$ Mpc. Finally,
figure 22 shows the results for cumulative number counting vs. redshift
produced by the model under consideration.

\section{Discussion}

\subsection{A Metric for a Smoothed-out and Averaged Fractal}

In the previous section it was shown that Tolman fractal solutions
do exist, and that the only ones compatible with observations are the
hyperbolic type solutions obtained by means of the specializations given
by equations (\ref{t29}) and (\ref{t30}).\footnote{ \ Notice however that
fractal solutions of elliptic and parabolic types may, in principle, be
obtained from other more complex specializations of the arbitrary
functions than the ones considered in this paper, and these solutions
could be compatible with observations.} As in equations (\ref{t29})
the function $\beta(r)$ is not constant, this means that the model has no
simultaneous big bang. In other words, in a model of this sort the big
bang singularity hypersurface occurred at different proper times in
different locations, and the age of the universe is different for
different observers at different radial coordinates. More specifically,
as $\beta(r)$ in equations (\ref{t29}) is an increasing function,
regions at smaller $r$ are younger than at bigger $r$, and the youngest
region of the model is ``here'', at $r=0$.

An universe model where some regions are older than others is not
as odd in terms of accepted ideas of galaxy formation as it might seem
at first. Inasmuch as the observed universe is lumpy, in a spatially
homogeneous Friedmann universe where $\rho = \rho (t)$ and the big bang
singularity is simultaneous, there must be density
fluctuations $\delta \rho / \rho$ of some kind in order to form
galaxies, and it is necessary to have some sort of metric perturbations
for that to happen. So, at the era of galaxy formation, which may be
defined as a hypersurface of constant time in order to agree with our
intuition in an unperturbed metric, in the perturbed metric the overdensities
$\rho_{\rm o}$ occur at time $t_{\rm o}$ and the underdensities
$\rho_{\rm u}$ occur at $t_{\rm u}$, and $t_{\rm o}$ must be different
from $t_{\rm u}$, otherwise there would not be any fluctuation at all.
In other words, due to the density fluctuations $\delta \rho / \rho$, in
the perturbed metric a hypersurface of constant density no longer coincides
with a hypersurface of constant time. Therefore, a deviation of the
Friedmann metric from spatial homogeneity, even if it is small, is
essential for lumpiness and, hence, some regions will inevitably have
different local times than others. In the perturbed metric we may even
define the era of galaxy formation as being a hypersurface of constant
density. In other words, a non-simultaneous big bang seems inevitable in
order to form galaxies in the standard scenario, even if those
differences in local times are small.

Note that this discussion assumes that a hypersurface of simultaneity is
defined by a specific value of the proper time, which is a logical thing
to do in an unperturbed metric. In a perturbed metric, however, one
could define the hypersurfaces of simultaneity in a different way, which
would mean a different choice of gauge by which the perturbed and the
non-perturbed spacetimes are related.

Nevertheless, considering it is desirable that fractal models be as close
as possible to their Friedmann counterparts, and also considering
mathematical simplicity, I shall take the 
specializations given by equations (\ref{t30}) as the best modelling of
a relativistic hierarchical (fractal) cosmology by Tolman's spacetime.
Let us now write this model explicitly. Its metric is expressed as
\begin{equation}
  dS^2=dt^2-\left( \frac{{R'}^2}{\cosh^2 r} \right) dr^2 -
       R^2 (d \theta^2+\sin^2 \theta d \phi^2); \ \ \ r \ge 0, \ 
       R[r,t(r)] \ge 0.
  \label{t31}
\end{equation}
The Einstein field equation for this metric may be written as an energy
equation (Bondi 1947; paper I)
\begin{equation}
    \frac{\dot{R}^2}{2} - U(R) = E(r),
    \label{t32}
\end{equation}
where
\begin{equation}
    U(R) = \frac{\alpha r^p}{4R}
    \label{t33} 
\end{equation}
is the effective potential energy, and 
\begin{equation}
     E(r) = \frac{1}{2} \sinh^2 r
 \label{t34}
\end{equation} 
is the total energy within $r$. The solution of equation (\ref{t32}) is
\begin{equation}
  R = \frac{ \alpha r^p}{8 E(r)} \left( \cosh 2 \Theta - 1 \right),
  \label{t35}
\end{equation}
where $\Theta$ is given by
\begin{equation}
  4 \left[ t(r) + \beta_0 \right] { \left[ 2 E(r) \right] }^{3/2} 
   = \alpha r^p \left( \sinh 2 \Theta - 2 \Theta \right),
 \label{t36}
\end{equation}
$t(r)$ is the solution of the past radial null geodesic
\begin{equation}
 \frac{d t}{d r} = - \frac{R'}{ \cosh r},
  \label{tt3366}
\end{equation}
and the local density is expressed as
\begin{equation}
  \rho = \frac{4 p { \left[ E(r) \right] }^2}{ \pi \alpha r^{p+1} R' 
	 { \left( \cosh 2 \Theta - 1 \right) }^2 }.
 \label{tt3377}
\end{equation}

The essentially new physical feature of the model above is its single
difference from the open Friedmann one: the form of the function for the
gravitational mass. That 
is given by $F = \alpha r^p$, while in Friedmann this function
must be $F = b_2 \sinh^3 r$. Remembering that $\alpha = 10^{-4} - 10^{-5}$
and $p=1 - 2.5$, the fractal metric (\ref{t31}) appears to have a more rarefied
dust  than its Friedmann equivalent. As fractal models are characterized
by a power law nature of their average densities, with fractional
exponents smaller than 3, it is hardly surprising that such models have
a more rarefied distribution of mass.

\subsection{Evolution of the Fractal Model and Comparison with the
Spatially Homogeneous Case}

The constant $\beta_0$ in equation (\ref{t36}) gives the universal
big bang time if we define ``now'' as $t=0$, and this fact allows us to
study the evolution of the fractal model (\ref{t31}) through the easily
computable manner of simply varying $\beta_0$. This investigation
permits us to answer the question of whether or not the fractal features of
the metric (\ref{t31}) are present at different epochs. Bearing  this
point in mind, I carried out simulations for different values of
$\beta_0$ in the interval $0 \le r \le 0.07$, but keeping $\alpha =
10^{-4}$ and $p=1.4$ in all of them. I found out that the model under
consideration remains fractal in integrations where $\beta_0 =$ 1.5, 2,
2.5, 3.6, 4.5 and 6, with the resulting fractal dimensions being $D=$
1.5, 1.5, 1.5, 1.4, 1.4 and 1.4, respectively. These results show that
the metric (\ref{t31}) effectively models a fractal distribution of dust
at different epochs, with a remarkable constancy in $D$. All those
integrations had $r = 0.07$, the final value of the integrating
parameter, corresponding to $z \cong 0.07$, although the luminosity
distance varied  from $d_l \cong 110$ Mpc when $\beta_0 = 1.5$ to $d_l
\cong 450$ Mpc when $\beta_0 = 6$. This variation may be physically
explained as due to changes in the Hubble constant itself, whose
values get bigger at earlier epochs. Finally, those results together
with some simulations on different values of $p$ suggest it is possible
to propose a simple, but very restricted, relationship between $p$ and $D$.
For $0 \le r \le 0.07$, $1.5 \le  \beta_0 \le 6$ and $ 1.4 \le p \le
2.5$ we can say approximately that $D = p \pm 0.1$.

An interesting question about the model (\ref{t31}) is to see how
it would compare with the spatially homogeneous open Friedmann one.
Looking at the results of the latter in figure 6 and the former in
figure 20 we can qualitatively see that the absolute value of the
difference in $\log \rho_v$ between the two models starts as zero, but
increases rapidly. In the analytical study of the Einstein-de Sitter
model presented in paper II it was shown that at the big bang
singularity hypersurface the volume density $\rho_v$ vanishes and the
luminosity distance $d_l$ goes to infinity, and this effect is a
consequence of the definition (\ref{t10}) of the volume density: at the
big bang the volume (\ref{t9}) is infinity, but the total mass is
finite. As already said at the end of \S 4, we can therefore expect
a similar asymptotic effect in both the
open Friedmann case and the fractal model (\ref{t31}), and this means
that the difference in $\rho_v$ between these two models should start
decreasing after reaching a maximum in its increase. Actually, figure 6
already shows a sharp decrease of $\rho_v$ in the open Friedmann case, but as
the integrations of the model (\ref{t31}) in figure 20 did not go far
enough in the past, the results cannot show where this maximum might be.
Nevertheless, based on this reasoning we can deduce that the
observational relations of the fractal model (\ref{t31}) appear to be
asymptotically Friedmann when calculated along the past light cone.

\subsection{Relations to Homothetic Self-Similarity}

Another topic which deserves some investigation is the relation, if
any, between self-similarity due to fractals and self-similarity due to
homothetic Killing vectors (Cahill \& Taub 1971). There is some
interest in this point because some attempts have
been made to explain large-scale voids and clusters by self-similar
perturbations of a Friedmann universe (Carr \& Yahil 1990), and also
because the first attempt to propose a workable relativistic
hierarchical cosmology was done assuming a homothetic self-similar
metric (Wesson 1978, 1979).

In general relativity a spacetime is called self-similar if all
metric components can be put in a form in which they are functions of a
single independent dimensionless variable which is a combination of the
spacetime coordinates. Mathematically, this corresponds to the existence
of homothetic Killing vectors, meaning that spherically symmetric
similarity solutions are unchanged by coordinate transformations of the
form $t \rightarrow bt$, $r \rightarrow br$, for any constant $b$
obeying the conformal transformation $g_{\mu \nu}(r,t) \rightarrow
\frac{1}{b^2} g_{\mu \nu}(r,t)$ (Cahill \& Taub 1971). Physically, spherically
symmetric similarity spacetimes with $\Lambda = 0$ contain no
fundamental scales or dimensional constraints (Henriksen \& Wesson
1978), and it would seem that these features make homothetic
self-similarity a possible mathematical version of the scaling idea
behind the empirical hierarchical clustering concept. However,
homotheties, as defined in general relativitity, are basically
geometrical features which will not necessarily translate themselves in
observable quantities. In other words, the geometrical scaling features
of the model do not necessarily mean that its observational relations
are also scaling. For this reason it seems that homothetic
self-similarity provides an unsatisfactory manner of modelling hierarchy.

It is beyond the aims of this work to make a general discussion
about the possible relations between these two types of self-similarity.
However it is of interest to investigate whether or not the fractal
solutions presented in \S 5 are homothetic. Recently Lemos \&
Lynden-Bell (1989) and Ponce de Leon
(1991) studied homotheties in Tolman models and found that specific
criteria are necessary to maintain the assumed similarity symmetry
in the solutions. For models where $\Lambda = 0$ (our case here),
they showed that the first criterion is for the function
$f(r)$ to be constant, that is, each comoving shell must have the
same total energy.
That immediately tells us that all fractal solutions of elliptic and
hyperbolic type studied here, including the metric (\ref{t31}), are
not homothetic, leaving only the parabolic solutions still to investigate.
The next criterion says that homothetic solutions with $f(r) = 1$
restrict the mass distribution to have the form $F(r) = \mbox{constant}
\times r^{2 \wp +3}$, where $\wp$ is a constant, and this is the case in
the solutions (\ref{t24}) and (\ref{t25}). The final requirement
for homothetic symmetry to hold in our Tolman solutions demands that
for $f=1$ and $\wp \not= 0$ the big bang hypersurface must be
of the form $\beta(r)= a + b r^{-\wp}$, where $a$ and $b$ are constants.
The solution (\ref{t24}) satisfies these three requirements only if
$p=3-2q$. The solution (\ref{t25}) can have $\beta(r)$ reduced to its
form in equations (\ref{t24}), but then the value $p=1$ must hold to
satisfy the third requirement. Therefore, very restricted cases of the
fractal solutions (\ref{t24}) and (\ref{t25}) have also homothetic
self-similarity.

In conclusion, from what we have seen it does appear valid to say
that fractal self-similarity is a much weaker requirement on the
solutions than homotheties, and although we have reached this conclusion
looking in more detail only at the Tolman spacetime, based on the
self-similar requirements for both cases it seems reasonable to suppose
that this conclusion may well be valid in general.

\subsection{Fitting Condition}

In the previous section of this paper we dealt with the problem of
finding a specific Tolman model which best represents the observed
inhomogeneous distribution of galaxies, and in that respect it was
concluded that the metric (\ref{t31}) is the simplest one to achieve
this aim. In other words, what was being sought was the optimal way of
fitting the Tolman metric to the real lumpy large-scale structure of the
universe. It was discussed in paper I that it may be desirable for us to
have a Friedmann metric as background spacetime to the inhomogeneous
Tolman region (or regions) used here to describe a fractal distribution of
galaxies, and having found the specific forms for this inhomogeneous
region an important question arises at once: what are the implications
that the fractal metric (\ref{t31}) and the hyperbolic solution
(\ref{t29}) bring to a possible Friedmann spacetime background?
Answering this question is equivalent to finding a response to the
``fitting problem'' in cosmology, in the specific context of this work.

Ellis \& Stoeger (1987) have outlined the fitting problem as
being the search for an ideal Friedmann model which best fits another
cosmological model that gives a realistic representation of the
universe, including all inhomogeneities down to some specified length
scale. In Ellis \& Stoeger's words, ``the approach resembles that used
in geodesy, where a perfect sphere is fitted to the pear-shaped earth;
deviations of the real earth from the idealised model can then be
measured and characterized''. Various ways in which this approach might
be tried are discussed in detail by them, however here we shall restrict
ourselves to the specific one outlined by Ellis \& Jaklitsch (1989)
where the matching between the Tolman and Friedmann spacetimes is
interpreted as a fitting condition.

It was shown in paper I that the Darmois junction conditions between
the two spacetimes under consideration require that $f=g'$ on the
joining surface. Here $g= \sin r$, $r$, $\sinh r$ is the function which
determines the curvature of the Friedmann model. As both the solution
(\ref{t29}) and the metric (\ref{t31}) are of hyperbolic type with $f=
\cosh r$, there is no way of satisfying this condition when $r > 0$
unless we have $g = \sinh r$. Therefore, the first response to the
fitting problem in this context says that our Tolman fractal solutions
imply an open Friedmann background model.

It was also shown in paper I that the matching between these two
spacetimes severely restricts the gravitational mass inside the Tolman
cavity. If $m(r) = F(r)/4$ is the gravitational mass of the Tolman
region within a comoving radius $r$, and $\overline{m}(x) = 4 \pi \mu
a^3(t) g^3(x)/3$ is its Friedmann equivalent for a radius $x$ and dust
density $\mu$, the junction conditions demand 
\begin{equation}
 m ( \Sigma_0 ) = \overline{m} ( \Sigma_0 ),
 \label{ultimaIII}
\end{equation}
where $\Sigma_0$ is the constant that defines the joining surface
$r=x=\Sigma_0$ between the two spacetimes. Therefore, as discussed by
Ellis \& Jaklitsch (1989), equation (\ref{ultimaIII}) allows us to
choose the Friedmann background model whose density is appropriate to
our lumpy Tolman model. Let us see in more detail how this background
spacetime can be specified.

In the open Friedmann model the local density is given by
\[
  \mu = \frac{3 b_2}{16 \pi a^3(t)},
\]
and as $F = \alpha r^p$ in our fractal models, we can then write the
equation (\ref{ultimaIII}) as
\begin{equation}
 \alpha {\Sigma_0}^p = b_2 \sinh ^3 \Sigma_0.
 \label{ultimaIV}
\end{equation}
The value of the parameter $b_2$ is what we are seeking in order to
determine precisely the Friedmann background and give a more accurate
answer to the fitting problem in this context. Equation
(\ref{ultimaIV}) shows that $b_2$ is dependent on the other three
parameters $\alpha$, $p$, $\Sigma_0$, and hence, there is a certain
degree of flexibility in choosing the mass of the background spacetime. Thus,
even when the interior region is determined by known values of $\alpha$
and $p$, different values of $b_2$ are obtained according to exactly
where the joining surface is located.

We can work out how the Friedmann background is in the case of the
numerical integrations of the model (\ref{t30}) shown in figure 20. We
have $\alpha = 10^{-4}$, $p=1.4$, and if we take $\Sigma_0 = 0.07$ (the
value where the numerical evaluation ends) we get $b_2 = 0.007$,
$\Omega_0 \cong 0.002$ and $H_0 = 83$ km/s/Mpc. This low value for
$\Omega_0$ is in the lower limits of the interval where it has been
reportedly measured. However, it is important to notice that in the
approach used in this work no kind of dark matter was considered, but
only the luminous matter associated with galaxies. Galactic luminous
matter gives a value for $\Omega_0$ of the same order of magnitude as
the one found for the background model above (see White 1990, p. 38).

As a final remark, we have so far considered the interior Tolman
region joining directly to the exterior Friedmann metric. That does not
need to be always the case and we can envisage an interior region
surrounded by one or more intermediary regions with higher or lower
densities, in a scheme designed to model specific observational
features. For example, a structure like the ``Great Wall'' (Geller \&
Huchra 1989; Ramella, Geller \& Huchra 1992) could be modelled by an
intermediary overdensity region before the background spacetime is
reached, and with an underdensity interior fractal Tolman region 
(see Bonnor \& Chamorro 1991 on how to join an underdensity Tolman
region to an overdensity intermediary section). Such a modelling will
obviously increase the value of $\Omega_0$ for the fitted Friedmann
background. This method, however, demands more detailed work in
order to achieve a model where this sort of structure is precisely
characterized. I shall not pursue further this study here.

\subsection{Criticism of Criticisms of Fractal Cosmology}

As the final issue of this section, I shall discuss some of the
objections raised by some authors against a fractal cosmology. The first
type of criticism is contrary to a possible unlimited fractal pattern
for the large-scale clustering of galaxies, and although some of the
critical voices do accept fractals at small scales, their objections are
usually based either on reasoning from the 2-point angular correlation
function (Peebles 1989), or on supposed strong theoretical limitations
of the standard Friedmannian cosmology. Therefore, in one way or another
those authors see the strong need for a crossover to homogeneity on the
fractal structure, at a scale yet to be agreed upon.

Criticisms based on the angular correlation function have been
addressed by Coleman \& Pietronero (1992) and I shall not discuss them
here, although this point was briefly mentioned in paper II. The
theoretical criticisms, on the other hand, must be addressed in this
paper, and for this purpose I shall reproduce  here two quotations from
Mart\'{\i}nez (1991) which well represent this point of view,
although he is by no means the only one to raise such kind of
objections. Mart\'{\i}nez  states that ``...~it should be noted that in
the standard cosmology, the distribution of mass must tend to a non-zero
finite density when averaged over large volumes'' \footnote{ \ The expression
``large volumes'' used in this quotation is imprecise, and can be
interpreted as meaning either big local volumes or volumes which are big
enough to be no longer considered as local. In a cosmological context the
latter is more appropriate and from now on I shall assume the expression
``large volumes'' to mean non-local ones.}. From a relativistic
point of view, the problem with this statement is its failure to specify
where this average is supposed to be carried out. It is correct to say
that if in a Friedmannian cosmology we make averages of density at
spacelike hypersurfaces of constant $t$, those averages cannot be zero
as this cosmology is spatially homogeneous. However, at large volumes
such averages would be observationally irrelevant as astronomy in the
electromagnetic spectrum is actually made along the backward null cone
and {\it not} at such spacelike hypersurfaces. Therefore, the statement
above is only true in an unobservable situation, and considering that
voids and clusters of galaxies were, and still are being, identified in the
so-called redshift space, which lies on the past null cone, this is
where the average of density must be carried out. Hence, the quotation
above is inappropriate as an objection to a fractal structure for
the distribution of galaxies. These two types of averages will only coincide
locally, and it will depend on the model and the value assumed for the
Hubble constant in order to establish what scales are local, although,
in any case they will certainly differ at large volumes. In effect, in
this imprecise formulation this statement may actually reinforce, or be
taken by, a common misinterpretation of the standard model, which, due to
inappropriate Newtonian analogies, confuses the model's geometry with
its observable quantities .

Mart\'{\i}nez (1991) goes on and states that ``... a fractal
universe without a crossover to homogeneity (...) implies a vanishing
density for very large volumes and this idea cannot be accepted without
creating important additional problems''. It was shown in paper II (and
rediscussed in \S 4) that
if averages on density are made along the backward null cone, at the big
bang singularity hypersurface the luminosity distance goes infinite and
this average density is zero. That happens in the Einstein-de Sitter
model, the most popular of the cosmological models, and no additional
hypothesis or change in the metric was done to achieve this result. The
point is, once more, where the average is made and which definition of
density is adopted. Thus, again the
statement is in fact an untenable objection since the standard cosmology
does have a vanishing average density without any important additional
problem. Having or not having zero average density in the model is just
a question of interpretation.

It should be clearly understood that the above criticisms of
Mart\'{\i}nez's (1991) statements are  made solely on the grounds of the
standard Friedmannian cosmology, and there is no need whatsoever to mention
any fractal hypothesis in order to show the impreciseness and
inappropriateness of such statements. The important point being that
even accepting the spatially homogeneous standard Friedmannian cosmology,
this model tells us we would only be able to see, through our telescopes,
its homogeneity locally. 

Closely related to this point of local homogeneity is the issue, one
could argue, of how we would understand in this context the reported
uniform distribution of some deep samples like radio sources. In the
first place it must be made very clear that the discussion made so far
is aimed at showing that from a theoretical point of view there is no
constraint to an unlimited fractal distribution, even from within a
Friedmannian framework, but that does not imply the fractal system
is indeed limitless, as an upper cutoff to homogeneity is not yet ruled
out. \footnote{ \ The proposal for an upper cutoff to homogeneity in the
fractal system appears to have been initially advanced by Ruffini, Song
\& Taraglio (1988), although Pietronero (1987) had already made a
discussion about this issue.}
Nonetheless, a simple calculation in the Einstein-de Sitter model, as
presented in paper II, will show that a 30\% decrease in $\rho_v$ (from 
the value at present time $t=0$) occurs at $z~\approx~0.1$ ($d_l~\approx~500$ 
Mpc), and this means that even considering such high error in the 
determination of the volume density, this is, roughly speaking, the maximum
approximate range where the homogeneity of this model could be
observed. Beyond this range the homogeneity of the Einstein-de Sitter
model would no longer be observed in the past light cone. Thus the first issue
raised by this result is a problem of methodology: curvature effects occur
in Friedmann models at much closer ranges than usually assumed, and
this means that those surveys must consider in their data analysis
expressions along the past light cone. Currently, calculations of
observational relations where
the backward null cone is taken into account is a very much neglected
problem in cosmology. Secondly, if it is confirmed that in those deep 
surveys the distribution is really uniform, that would put the
Friedmann model in even greater difficulties as it would appear to
predict inhomogeneity in deeper ranges where this would not be observed.
Thirdly, the sceptical viewpoint on this issue would be to argue
that usually deeper observations are less precise than shallower ones,
and previous claims of the so-called homogeneous ``fair-sample'' finally
being observed did not stand once more refined and complete observations
were made. Historically, the range at where the homogeneity is, or would
be, finally reached has being pushed further and further away as more
complete data become available and observational techniques improve, and so,
the sceptic may say, we may not have necessarily seen the end of this story.

In addition to the points discussed above, a second kind of criticism
to the fractal cosmology has
been voiced by Peebles, Schramm, Turner \& Kron (1991) in the following
form. ``If the galaxy distribution had been observed to follow a pure
scale-invariant fractal, (...) the closely thermal spectrum and isotropy
of the cosmic background radiation in this highly inhomogeneous Universe
would have been a deep puzzle''. First of all, it should be said that in an
inhomogeneous model with a Friedmann background, the apparent discrepancy
between the inhomogeneity of the model and the isotropy of the microwave
background is not really an issue as the junction conditions already 
require that an overdensity must be compensated by an underdensity before
the uniform region is reached, in order that the average densities will
be the same (see paper~I, \S 4). Nevertheless, the most important point
is the result already obtained in paper II and extended in this paper for the
other Friedmann models: the standard Friedmannian cosmology may be taken
to be inhomogeneous depending on how we look at it. That means that the
apparent contradiction between inhomogeneity and the isotropy of the
microwave background may not be a contradiction at all. This is an
essential point in order to put Peebles, Schramm, Turner \& Kron's 
statement above into perspective, as we have already seen in this paper
and in paper~II that at relatively modest luminosity distances and
redshifts, even the standard spatially homogeneous Friedmannian cosmology
becomes highly inhomogeneous on the past light cone because $\rho_v$ departs
considerably from its constant value in our constant time hypersurface the
further into the past we look, and $\rho$ is dependent on $r$. Considering
that so far even the deepest all-sky redshift surveys have failed to reach the
so-called ``fair sample'' where the homogeneity is supposed to be, waiting
for us to discover it, perhaps it is about time to ask if the cosmic
background radiation could be accommodated in a different cosmology,
or wonder whether it is really a deep puzzle.

%To conclude this section, we are left with an important point to consider
%after this discussion.
After all this discussion, we are left with an important point to
consider. If the supposed theoretical need for a crossover
to homogeneity in the fractal system is much weaker than previously
thought, we have the question: is it really necessary? Since we have
seen here and in paper II that even the Friedmann models do not seem to
remain homogeneous along the past null geodesic (see figures 1, 2, 6
and 7), if the homogeneity of the standard model does not survive, where
is the strong need for a crossover? In paper I it was assumed that the
Tolman metric would
eventually join a Friedmann background and, among other things, I argued
the need for that was to make the model compatible with a different
interpretation of the Copernican principle. However, in the light of
inhomogeneity even in the Friedmann metric, it could be argued that from
an observational point of view, that is, in calculations along our
past null cone, if the relativistic fractal cosmology developed in this
series of papers has or has not a Friedmann background might well be
irrelevant.

\section{Conclusion}

In this work I have presented numerical solutions of particular
Tolman models of elliptic, parabolic and hyperbolic types featuring
fractal behaviour along the past light cone. The initial conditions of
the numerical problem is discussed in detail and I found three
different sets of initial values that can, in principle, be used to
solve the problem numerically. In practice, however, I have made use
of only two due to strong stiff behaviour in the numerical solutions.
The algorithm used to get the solutions has been described, as well as
some details of the manner it was implemented in the computing code.

The spatially homogeneous Friedmann models were treated as
special cases of the Tolman solution and solved numerically. In an
extension of the results of paper~II, I have found that the $K = \pm 1$
Friedmann cases also do not appear to remain homogeneous along the
backward null cone, and this is explained
because the models are {\it spatially} homogeneous: when
the density is averaged along the null geodesic, going through
hypersurfaces of constant $t$, but with different values for the
local density, this average changes. Furthermore, as in the open and
closed Friedmann models the volume density decreases the farther we go,
it would seem that similarly to the Einstein-de Sitter model studied
in paper~II, all cases of the standard cosmological model have
{\it zero} global density at the big bang singularity hypersurface.

Fractal solutions were obtained in all classes of Tolman's
solution and the mathematical criterion for getting them was their
obedience to the de Vaucouleurs' density power law. Nonetheless,
considering that observationally realistic models have to follow the
linearity of the luminosity distance--redshift diagram, have a Hubble
constant within the currently accepted uncertainty and fractal
dimension in agreement with measurements of the 2-point correlation
function, I found only hyperbolic type solutions obeying all these
criteria.

The simplest fractal hyperbolic solution has simultaneous
big bang and it was found that this model has fractal features in many
different epochs, does not have homothetic self-similarity (although
some of the parabolic fractal solutions do) and if we assume a
Friedmann background spacetime, the matching between these two metrics
imply that the background must be an open Friedmann model. A different
scenario where the ``Great Wall'' could be modelled was also
considered. The paper ends with a discussion on some objections raised
by some authors against a fractal cosmology, where I have shown how some
are imprecise and inappropriate, leading to untenable objections to
this sort of cosmology.

And what about the origin of this fractal system? If a Friedmann
background is assumed, it might be a cellular-type structure
(Feng, Mo \& Ruffini 1991; Fabbri \& Ruffini 1992 and references
therein), but with the difference that an upper cutoff to homogeneity
in the past light cone is not compulsory, as discussed in
the previous section. In this case this structure could be a result of
a fragmentation process (Ruffini, Song \& Taraglio 1988). If we do not
assume a Friedmann background, Bonnor (1974) showed that a hyperbolic
Tolman model can start inhomogeneous, remain inhomogeneous during the
course of its evolution and at large times, when the dust has already
escaped from its own gravitational field, there is nothing that could
alter the model's density distribution. In this case the function
$\beta (r)$ does not appear neither in the function $R(r,t)$ nor in
the density $\rho $, which was interpreted as meaning that at large
$t$ the model has ``forgotten'' its big bang, at least as far as the
density is concerned. In conclusion, the origin of the fractal
structure appears to be an open problem.

\vspace{15mm}
\begin{flushleft}
{\large \bf Acknowledgements}
\end{flushleft}
\vspace{5mm}

I thank M. A. H. MacCallum for his guidance during this research
and for valuable suggestions and discussions in all stages of this
work. I am also grateful to W. B. Bonnor for discussions and
suggestions, J. H. E. Cartwright for discussions about stiff problems
and T. Allen, A. B. Burd and J. E. F. Skea for suggestions on
numerical integrations. Finally I thank W. Seixas for his help with
computing problems and A. Koutras for discussions on homotheties.
This work had the financial support of the Brazilian Agency CAPES.

%\newpage            
\vspace{1cm}
\begin{flushleft}
{\large \bf References}
\end{flushleft}
  \begin{description}
    \item Bondi, H. 1947, M. N. R. A. S., 107, 410.
    \item Bonnor, W. B. 1974, M. N. R. A. S., 167, 55.
    \item Bonnor, W. B. \& Chamorro, A. 1990, Ap. J., 361, 21.
    \item Bonnor, W. B. \& Chamorro, A. 1991, Ap. J., 378, 461.
    \item Cahill, A. H. \& Taub, M. E. 1971, Commun. Math. Phys., 21, 1.
    \item Calzetti, D. \& Giavalisco, M. 1991, in Applying Fractals in
          Astronomy, ed. A. Heck \& J. M. Perdang, p. 119
	  (Springer-Verlag: Berlin).
    \item Carr, B. J. \& Yahil, A. 1990, Ap. J., 360, 330.
    \item Cartwright, J. H. E. \& Piro, O. 1992, Int. J. Bifurcation and
          Chaos, in press.
    \item Coleman, P. H. \& Pietronero, L. 1992, Phys. Reports, 213, 311.
    \item Coley, A. A. \& Tavakol, R. K. 1992, Gen. Rel. Grav., 24,
          835.
    \item Deng, Z.-G., Wen, Z., \& Liu, Y.-Z. 1988, in Large Scale
          Structures of the Universe, Proc. of the 130th IAU Symposium,
	  ed. J. Audouze, M.-C. Pelletan \& A. Szalay, p. 555 (Kluwer Academic
	  Publishers: Dordrecht).
    \item Ellis, G. F. R. \& Stoeger, W. 1987, Class. Quantum Grav., 4,
          1697.
    \item Ellis, G. F. R. \& Jaklitsch, M. J. 1989, Ap. J., 346, 601.
    \item Fabbri, R. \& Ruffini, R. 1992, Astron. Astrophys., 254, 7.
    \item Feng, L. L., Mo, H. J. \& Ruffini, R. 1991, Astron.
          Astrophys., 243, 283.
    \item Gear, C. W. 1971, Numerical Initial Value Problems in Ordinary
          Differential Equations (Prentice-Hall: Englewood Cliffs).
    \item Geller, M. J. \& Huchra, J. P. 1989, Science, 246, 897.
    \item Harrison, E. 1993, Ap. J., 403, 28.
    \item Henriksen, R. N. \& Wesson, P. S. 1978, Ap. Space Sci., 53, 429.
    \item Lambert, J. D. 1991, Numerical Methods for Ordinary
          Differential Systems (John Wiley \& Sons: Chichester).
    \item de Lapparent, V., Geller, M. J. \& Huchra, J. P. 1986,
          Ap. J. (Letters), 302, L1.
    \item Lemos, J. P. S. \& Lynden-Bell, D. 1989, M. N. R. A. S., 240, 317.
    \item Mart\'{\i}nez, V. J. 1991, in Applying Fractals in Astronomy,
          ed. A. Heck \& J. M. Perdang, p. 135 (Springer-Verlag: Berlin).
    \item Ortega, J. M. \& Poole, W. G., Jr. 1981, An Introduction to
          Numerical Methods for Differential Equations (Pitman:
	  Marshfield).
    \item Peacock, J. 1991, Nature, 352, 378.
    \item Peebles, P. J. E. 1989, Physica D, 38, 273.
    \item Peebles, P. J. E., Schramm, D. N., Turner, E. L. \& Kron, R. G.
	  1991, Nature, 352, 769.
    \item Pietronero, L. 1987, Physica, 144A, 257.
    \item Ponce de Leon, J. 1991, M. N. R. A. S., 250, 69.
    \item Press, W. H., Flannery, B. P., Teukolsky, S. A. \& Vetterling,
          W. T. 1986, Numerical Recipes: The Art of Scientific
	  Computing (Cambridge: Cambridge).
    \item Ramella, M., Geller, M. J. \& Huchra, J. P. 1992, Ap. J., 384,
          396.
    \item Ribeiro, M. B. 1992a, Ap. J., 388,  1 (paper I).
    \item Ribeiro, M. B. 1992b, Ap. J., 395, 29 (paper II).
    \item Ribeiro, M. B. 1992c, Ph.D. thesis, Queen Mary \& Westfield
          College - University of London.
    \item Ruffini, R., Song, D. J. \& Taraglio, S. 1988, Astron.
          Astrophys., 190, 1.
    \item Saunders, W., et al. 1991, Nature, 349, 32.
    \item Tolman, R. C. 1934, Proc. Nat. Acad. Sci. (Wash.), 20, 169.
    \item Wertz, J. R. 1970, Ph.D. thesis, University of Texas at Austin.
    \item Wesson, P. S. 1978, Ap. Space Sci., 54, 489.
    \item Wesson, P. S. 1979, Ap. J., 228, 647.
    \item White, S. D. M. 1990, Physical Cosmology, in Physics of the
          Early Universe, ed. J. A. Peacock, A. F. Heavens \& A. T.
	  Davies, p. 1 (Edinburgh University Press: Edinburgh).
  \end{description}
%\newpage
\vspace{1cm}
\begin{center}
 {\large Figure Captions}
 \vspace{10mm}
\end{center}
\begin{namelist}{Figurex10:x}
 \item[{\rm Figure 1:}] This graph is the same appearing in paper~II
      and shows the analytical results of a $\log~\rho_v$~vs.~$\log~d_l$
      plot for the Einstein-de Sitter model in the range
      $0.001~\leq~r~\leq 1.5$ and with $\beta_0=2.7$. The inhomogeneity
      of the model along the past null geodesic is clearly visible.
 \item[{\rm Figure 2:}] Plot of the numerical results of the volume
      density $\rho_v$  against the luminosity distance $d_l$ in the
      range $0 \leq r \leq 1.5$ for the Einstein-de Sitter model.
      Only steps bigger than $0.0025$ are shown here. Distance is given
      in Gpc ($10^9$ pc) and mass in units of $2.09 \times 10^{22}
      M_{\odot}$. By comparing these results with the analytical ones
      shown in figure 1 we can see that the numerical scheme produces
      the expected results.
 \item[{\rm Figure 3:}] Numerical integration of the null geodesic in
      the Einstein-de Sitter model. Time is in units of $3.26$ Gyr.
 \item[{\rm Figure 4:}] Numerical integration of the equation (\ref{t6})
      which gives the function $I(r)$ in the Einstein-de Sitter model.
 \item[{\rm Figure 5:}] The same results of figure 2 for the Einstein-de
      Sitter model, but with all steps visible. The numerical code takes an
      excessive number of steps around the origin to integrate this
      model, despite the smoothness of the integrated
      functions $t(r)$ and $I(r)$ shown in figures 3 and 4, respectively.
      The solution in this graph shows clearly the stiff behaviour of
      equations (\ref{t4}) and (\ref{t6}) at around $r=0$ in this model.
 \item[{\rm Figure 6:}] Numerical results of $\log \rho_v$ versus $\log
      d_l$ in an open Friedmann model. Very small steps are omitted and
      so stiffness is not visible. From now on the figures will not have
      visible stiff behaviour to avoid saturation.
 \item[{\rm Figure 7:}] Results of $\rho_v$ plotted against $d_l$ for the
      integrations of a recollapsing Friedmann model.
 \item[{\rm Figure 8:}] Numerical results of the parabolic model
      obtained with functions (\ref{t24}) integrated in the interval $0
      \leq r \leq 2$ and its straight line fitting carried out according
      to equation (\ref{u8}). The constants took the values
      $\alpha=10^{-5}, \ p=0.9, \ \beta_0=2, \ \eta_0=50, \ q=0.65$ and
      the resulting fitting coefficients are $a_1=-5.5$ and $a_2=-1.6$.
      Considering equation (\ref{t12}) the fitting coefficients tell us
      that the fractal dimension of the dust in this model is $D=1.4$
      and the lower cutoff constant is $\sigma=2.8 \times 10^6$.
 \item[{\rm Figure 9:}] Results of $\rho_v$ and $d_l$ for the
      integration of the parabolic model with functions (\ref{t25}) in
      the interval $0 \leq r \leq 5$.  The constants used had the
      values: $\alpha=10^{-4}, \ p=0.5, \ \beta_0=1.5, \ \eta_1=1000$.
      The fitted straight line is also visible and the resulting fitting
      coefficients are $a_1=-3.9$ and $a_2=-1.3$, which means a fractal
      dimension $D=1.7$ and a lower cutoff constant $\sigma=9.7 \times
      10^7$.
 \item[{\rm Figure 10:}] Distance-redshift diagram obtained by
      integrating the parabolic model of functions (\ref{t24}). Note the
      absence of a linear relation between the luminosity distance $d_l$
      and the redshift $z$. This fact makes this model observationally
      unrealistic.
 \item[{\rm Figure 11:}] Distance-redshift diagram obtained by
      integrating the parabolic model of functions (\ref{t25}). The
      diagram is only partially linear, but even so the Hubble constant
      associated with this part is just at or below the lower limit for
      $H_0$. This point implies that functions (\ref{t25}) are not a good 
      modelling for the observations.
 \item[{\rm Figure 12:}] Volume density $\rho_v$ vs. luminosity distance
      $d_l$ for the integration of the elliptic model given by functions
      (\ref{t26}) in the interval $0.001 \leq r \leq 0.07$. The constants
      took the values $\alpha=10, \ p=1.4, \ \beta_0=0.7, \ \eta_0=50,
      \ q=0.65$ and the resulting fitting coefficients are $a_1=-2.7$
      and $a_2=-1.3$, which means $D=1.7$ for this model.
 \item[{\rm Figure 13:}] Volume density $\rho_v$ vs. luminosity distance
      $d_l$ for the integration of the elliptic model given by
      functions (\ref{t27}) in the interval $0.001 \leq r \leq 0.07$. The
      constants used had the values: $\alpha=10, \ p=1.4, \ \beta_0=0.7,
      \ \eta_1=1000$. The fitting coefficients obtained are $a_1=-2.4$
      and $a_2=-1$. The models has fractal dimension $D=2$.
 \item[{\rm Figure 14:}] Distance-redshift diagram obtained by
      integrating the elliptic fractal model produced by functions
      (\ref{t26}). The diagram is not linear which makes the model
      incompatible with observations.
 \item[{\rm Figure 15:}] Distance-redshift diagram obtained by
      integrating the elliptic fractal model produced by functions
      (\ref{t27}). The diagram can be approximated to a linear relation,
      but that produces a too low value for $H_0$
 \item[{\rm Figure 16:}] Plot of the results of $\rho_v$ vs. $d_l$
      obtained by the numerical integration in the interval $0 \leq r
      \leq 0.07$ of the hyperbolic model constructed with functions
      (\ref{t28}). The values of the constants used to find these
      results are $\alpha=10^{-4}, \ p=1.9, \ \beta_0=3.6, \
      \eta_0=50, \ q=0.65$ and the fitting coefficients obtained
      are $a_1=-7.3$ and $a_2=-1.6$.
 \item[{\rm Figure 17:}] Distance-redshift diagram obtained with the
      hyperbolic model of functions (\ref{t28}). The diagram is
      approximately linear but the associated Hubble constant of about
      23 km/s/Mpc is too low comparing with the current uncertainty of
      $H_0$.
 \item[{\rm Figure 18:}] Plot of the results of $\rho_v$ vs. $d_l$
      obtained by the numerical integration in the interval $0 \leq r
      \leq 0.07$ of the hyperbolic model constructed with functions
      (\ref{t29}). The constants used to find these results are
      $\alpha=10^{-4}, \ p=1.4, \ \beta_0=3.6,\ \eta_1=1000$. The
      fitting coefficients obtained are $a_1=-6.2$ and $a_2=-1.7$. These
      coefficients mean a fractal dimension $D=1.3$ and a lower cutoff
      constant $\sigma=5.4 \times 10^5$.
 \item[{\rm Figure 19:}] Distance-redshift diagram obtained with the
      hyperbolic model consisted of functions (\ref{t29}). This is one
      of the best linear approximation for a $d_l$ vs. $z$ diagram and
      the slope gives $H_0 \cong 61$ km/s/Mpc, a value within the current
      uncertainty in the Hubble constant.
 \item[{\rm Figure 20:}] Results of the volume density $\rho_v$ vs.
      luminosity distance $d_l$ of the hyperbolic model (\ref{t30}) with
      simultaneous big bang. The integration is in the interval $0 \leq
      r \leq 0.07$ and the constants are $\alpha=10^{-4}, \ p=1.4, \
      \beta_0=3.6$. The fitting coefficients calculated are $a_1=-6.0$
      and $a_2=-1.6$, giving a fractal dimension $D=1.4$ and a lower
      cutoff constant $\sigma=8.7 \times 10^5$.
 \item[{\rm Figure 21:}] Distance-redshift diagram obtained with the
      hyperbolic model of functions (\ref{t30}). This is the best
      diagram  obtained for $d_l$ vs. $z$ and the slope  gives $H_0
      \cong 80$ km/s/Mpc, a value which is not only within the current
      uncertainty in the Hubble constant but also agrees with recent
      measurements on it (see Peacock 1991).
 \item[{\rm Figure 22:}] Plot of the results for the cumulative number
      counting $N_c$ vs. the redshift $z$ given by the integration of
      the model (\ref{t30}) with the same parameters as in figure 20.
\end{namelist}
\newpage
\begin{figure}
\includegraphics[width=12cm,angle=-90]{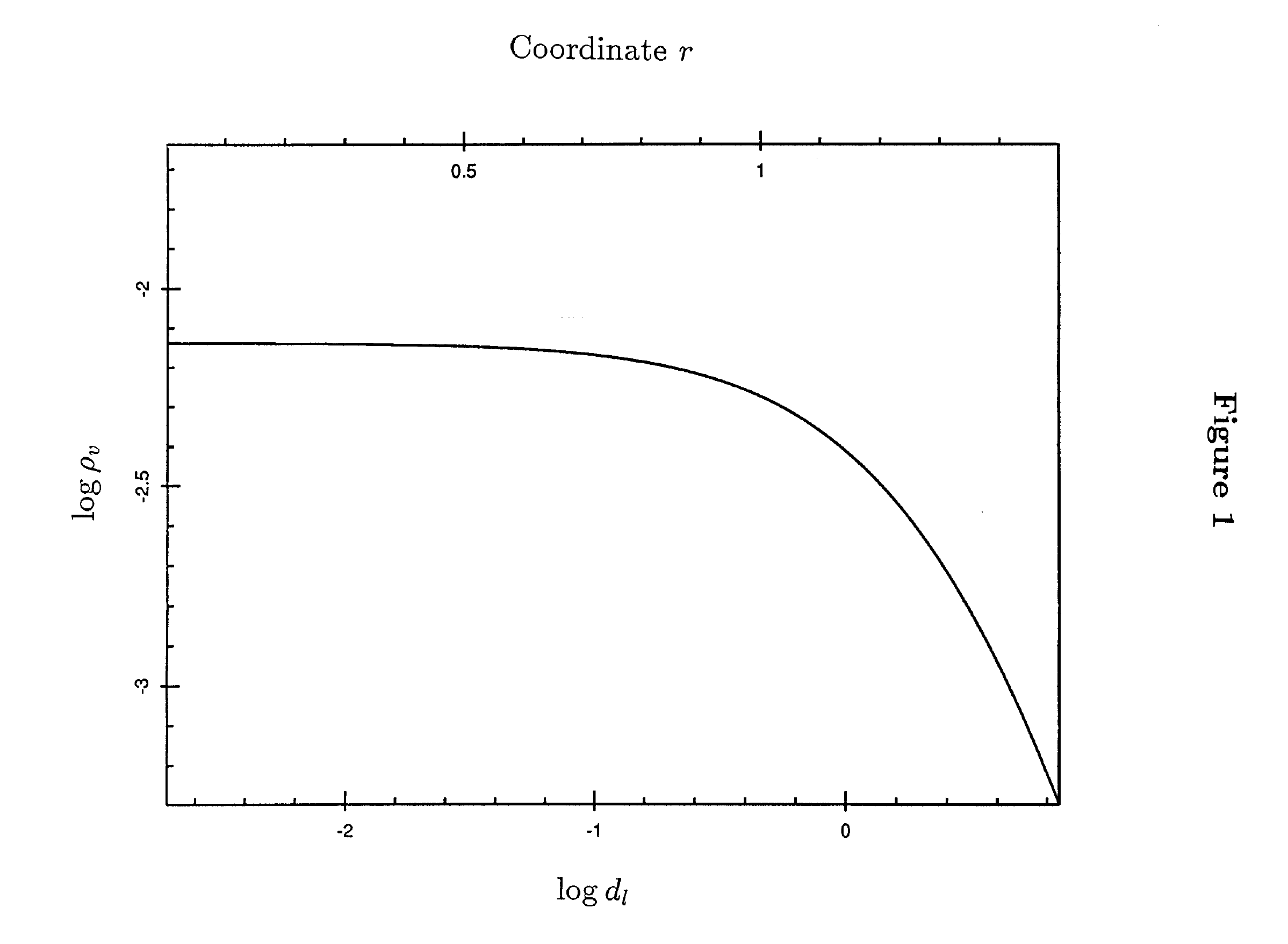}
\end{figure}
\begin{figure}
\includegraphics[height=18cm]{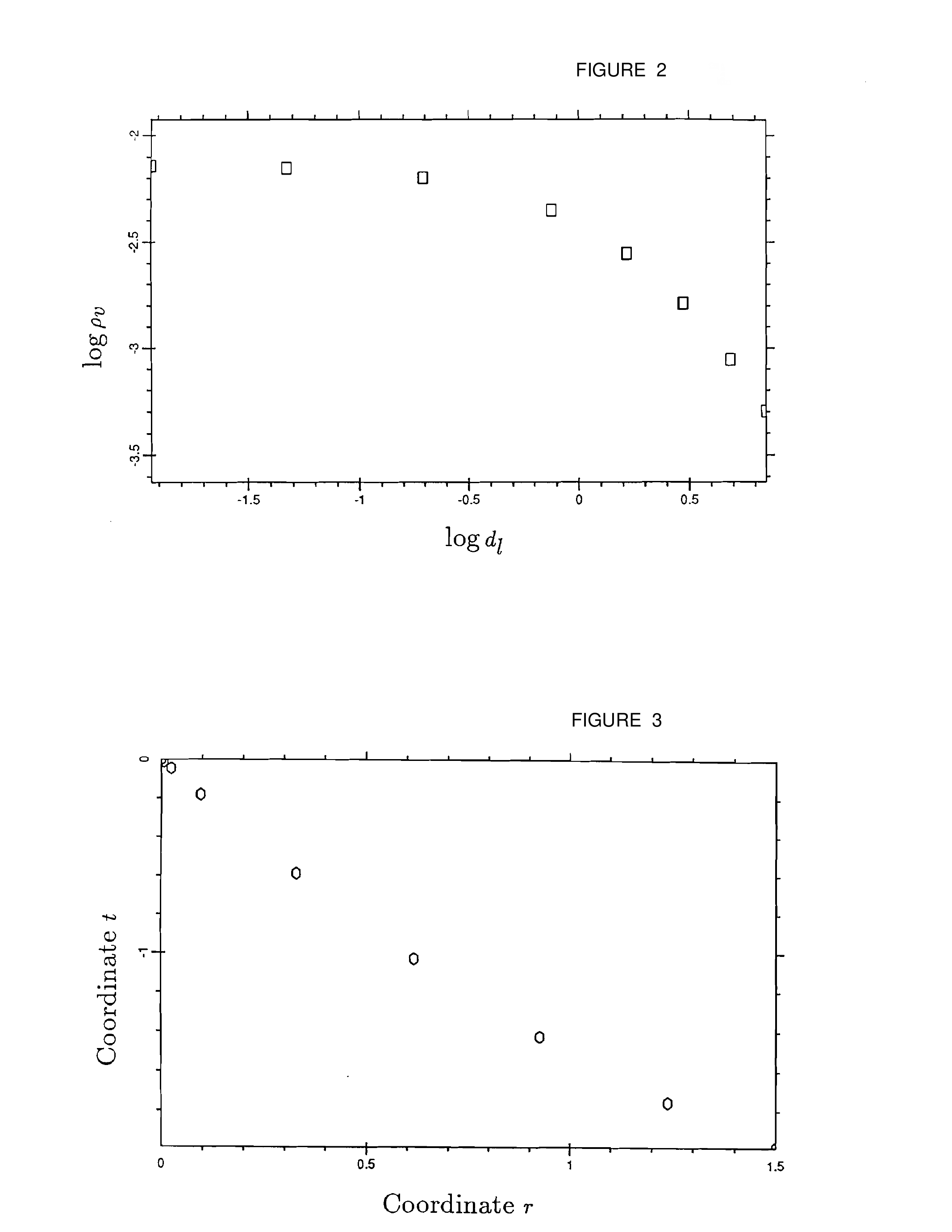}
\end{figure}
\begin{figure}
\includegraphics[height=18cm]{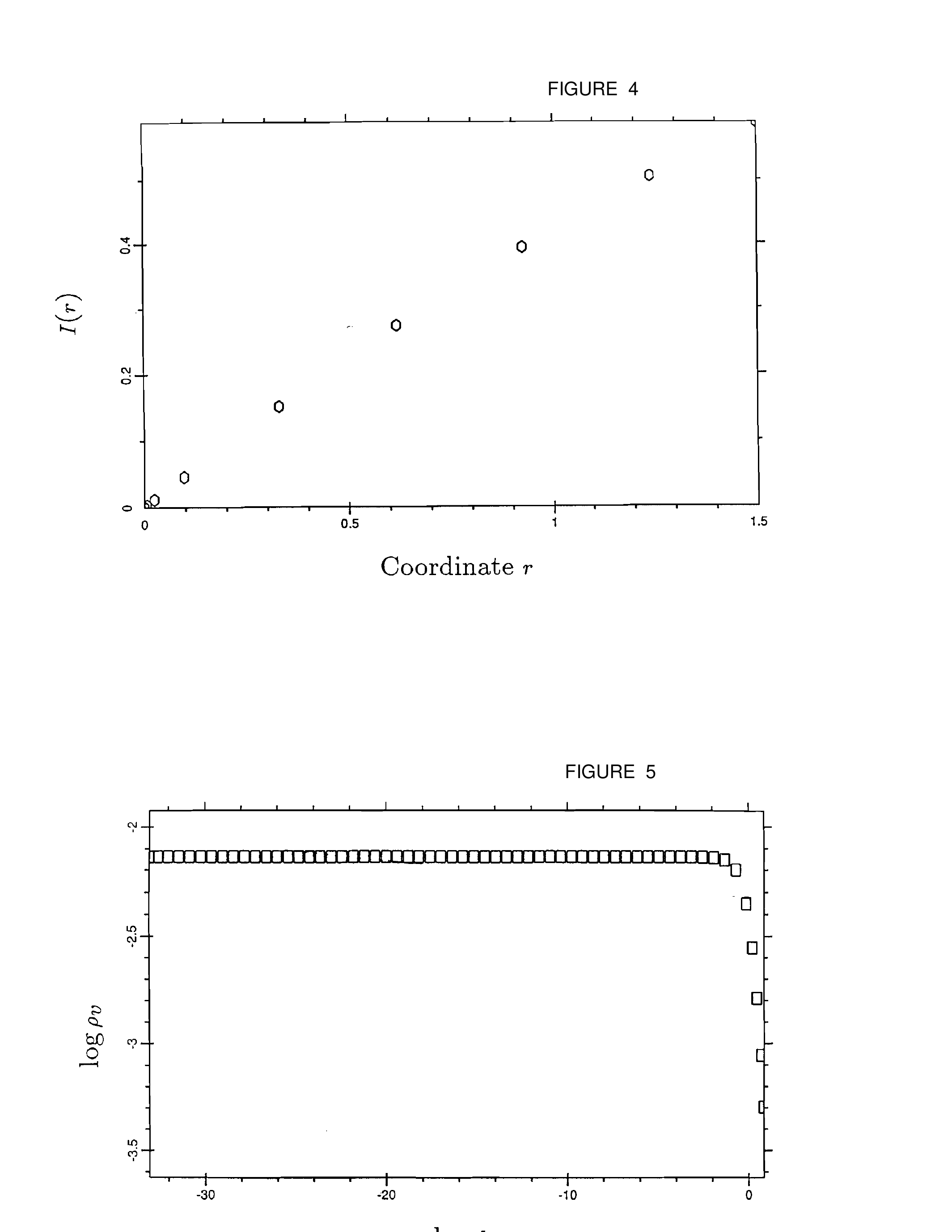}
\end{figure}
\begin{figure}
\includegraphics[height=18cm]{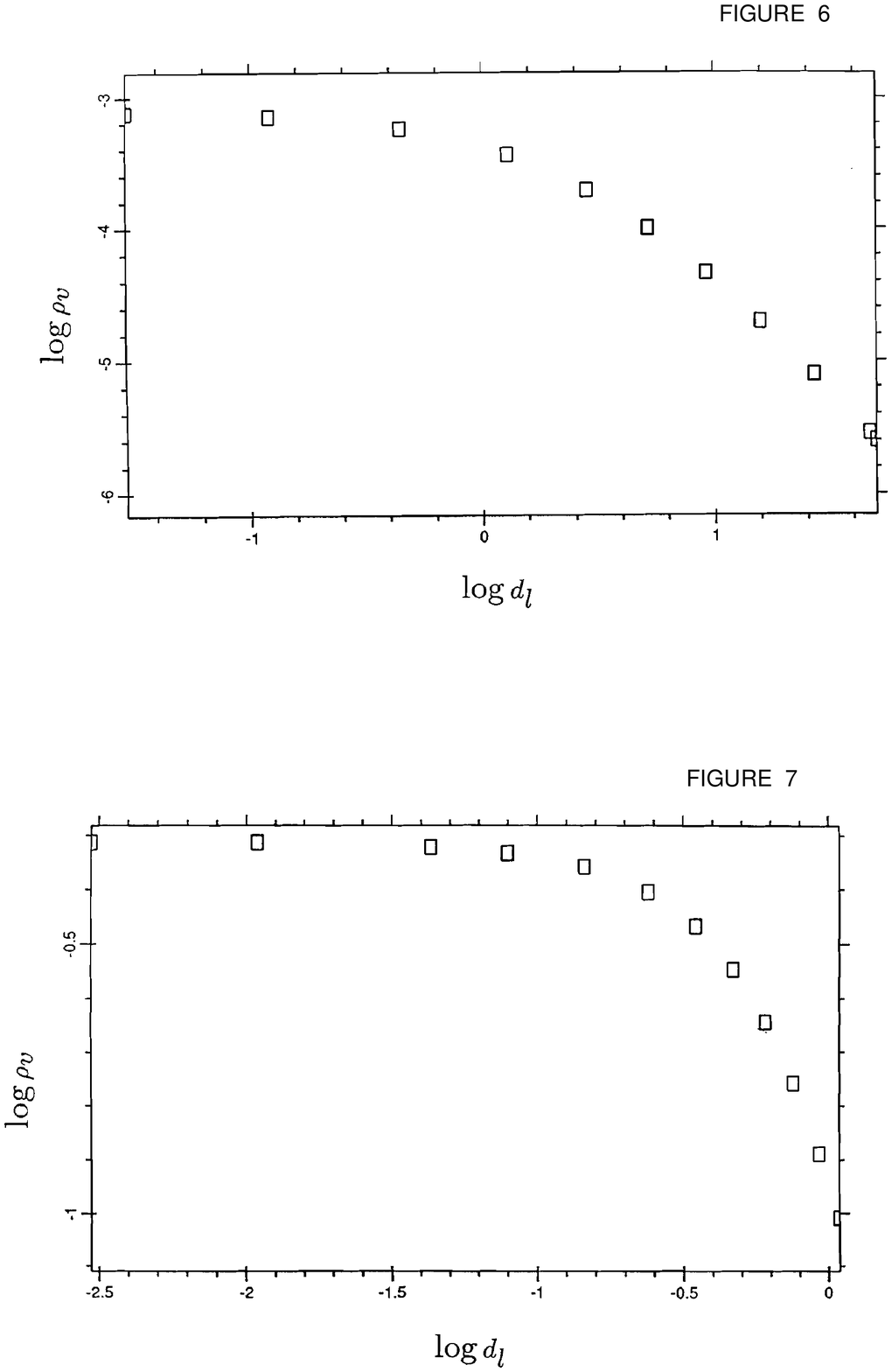}
\end{figure}
\begin{figure}
\includegraphics[height=18cm]{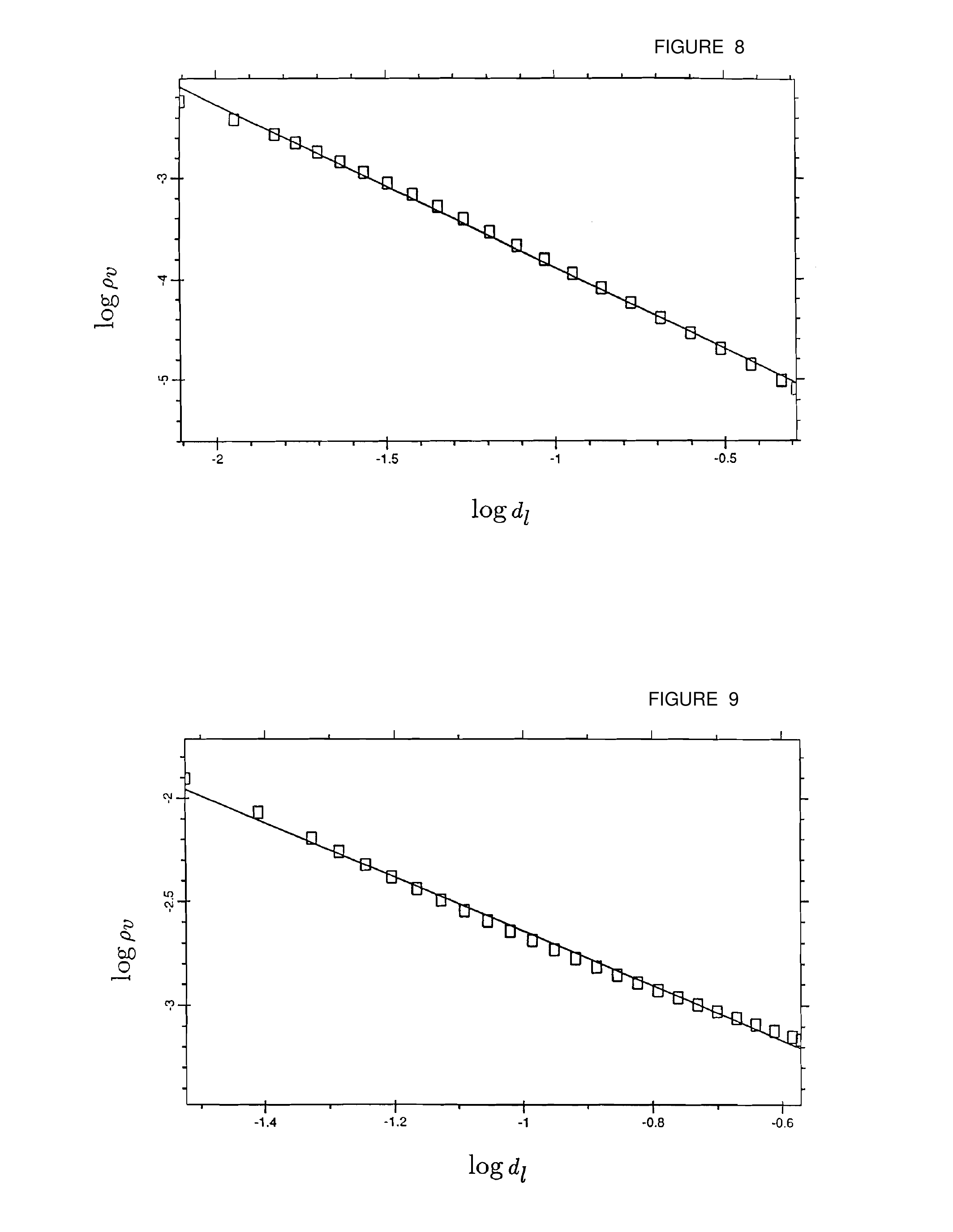}
\end{figure}
\begin{figure}
\includegraphics[height=18cm]{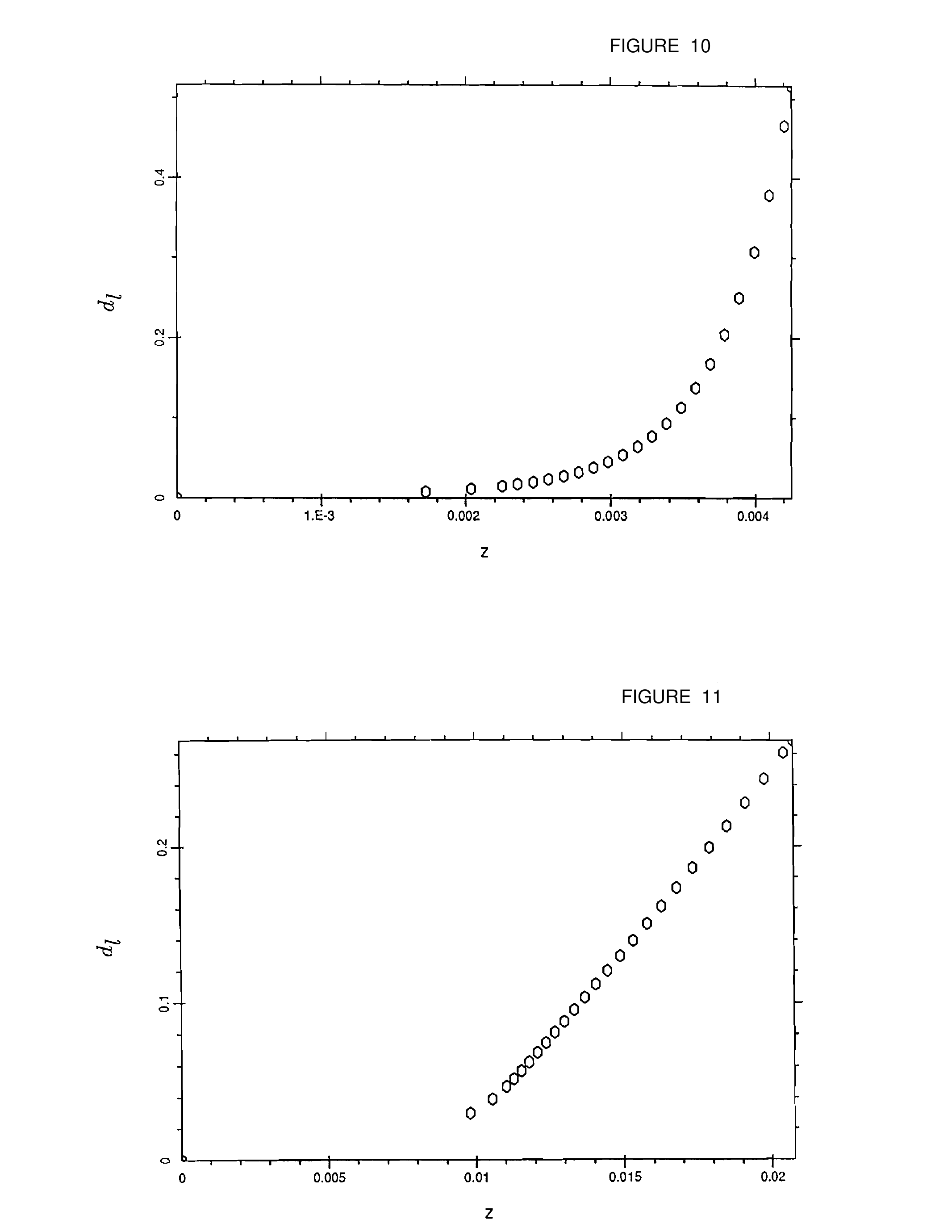}
\end{figure}
\begin{figure}
\includegraphics[height=18cm]{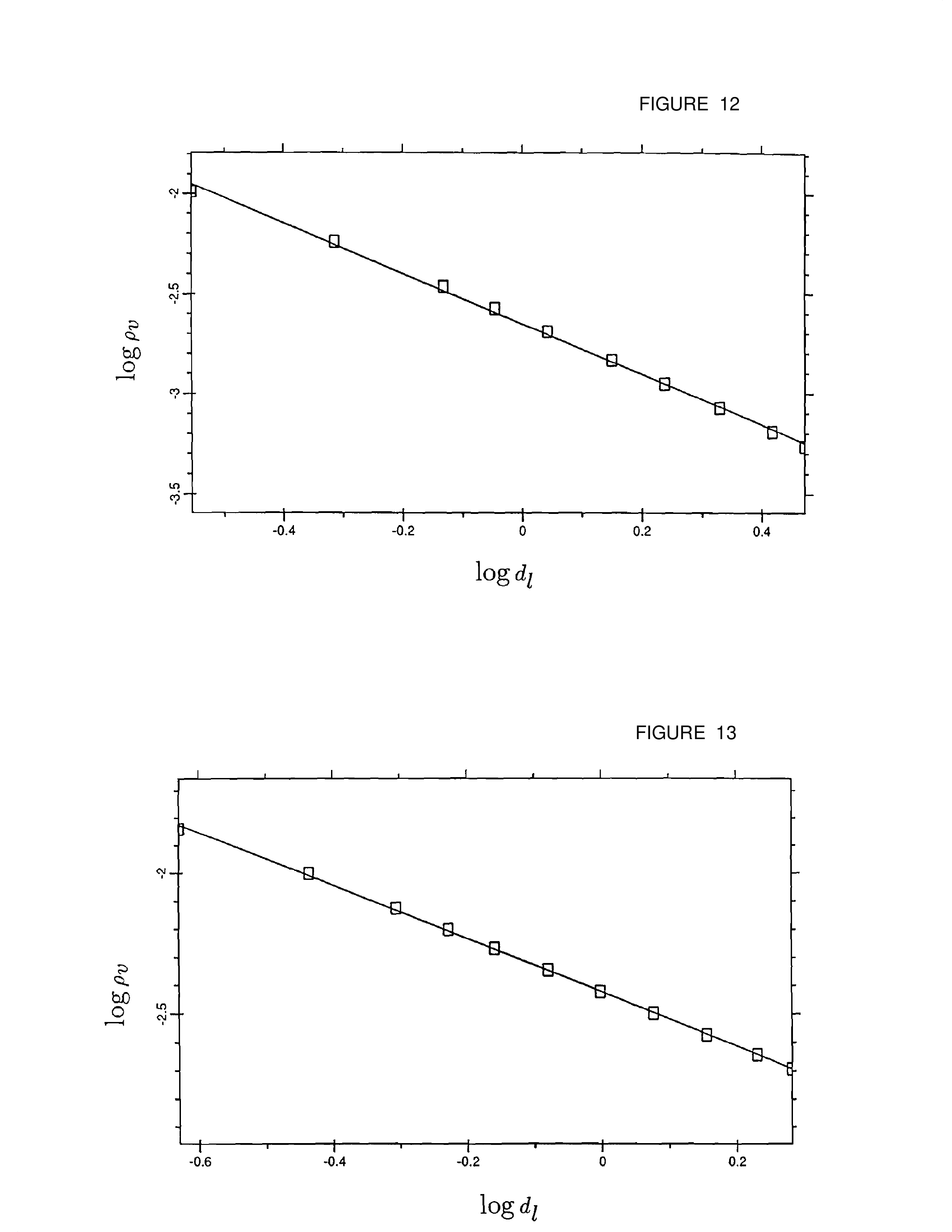}
\end{figure}
\begin{figure}
\includegraphics[height=18cm]{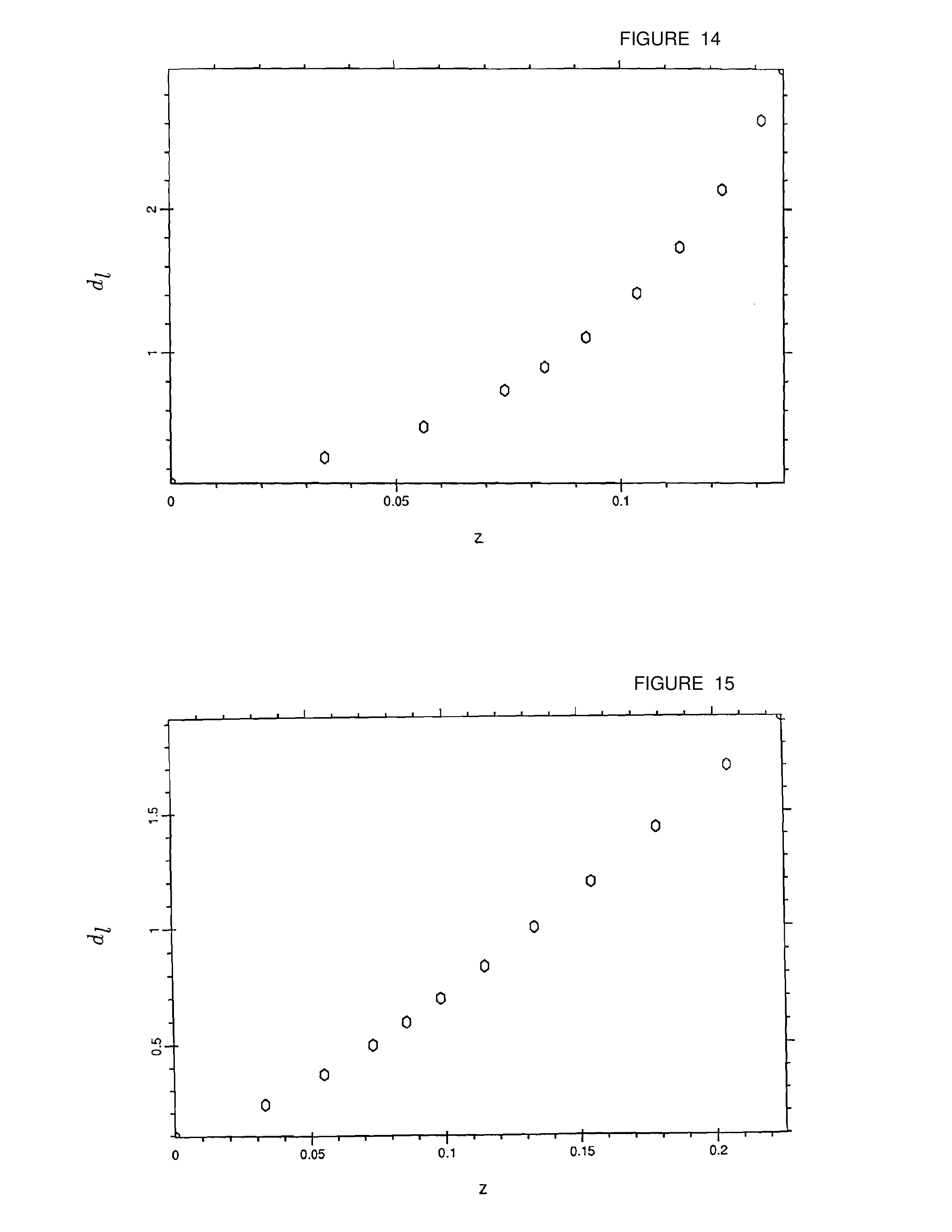}
\end{figure}
\begin{figure}
\includegraphics[height=18cm]{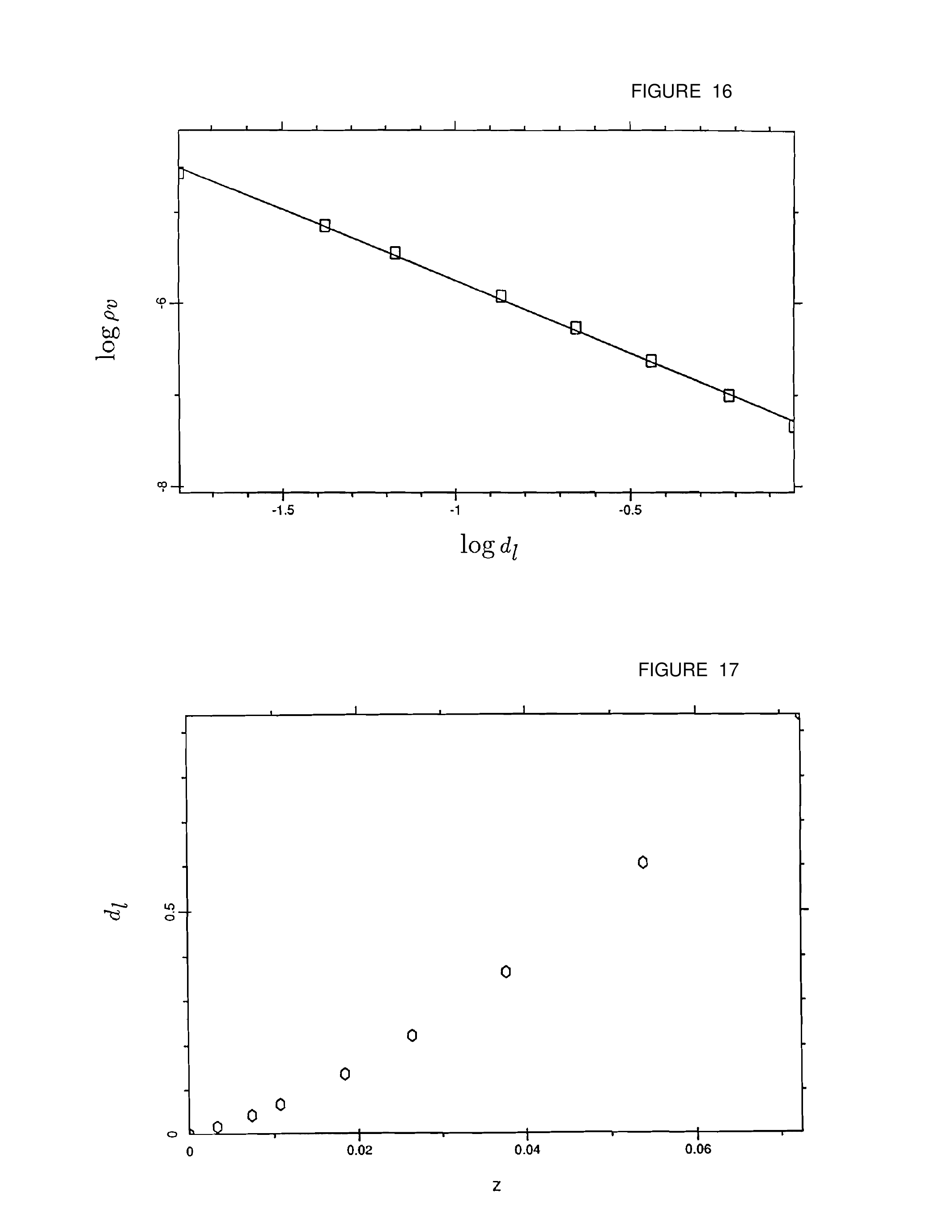}
\end{figure}
\begin{figure}
\includegraphics[height=18cm]{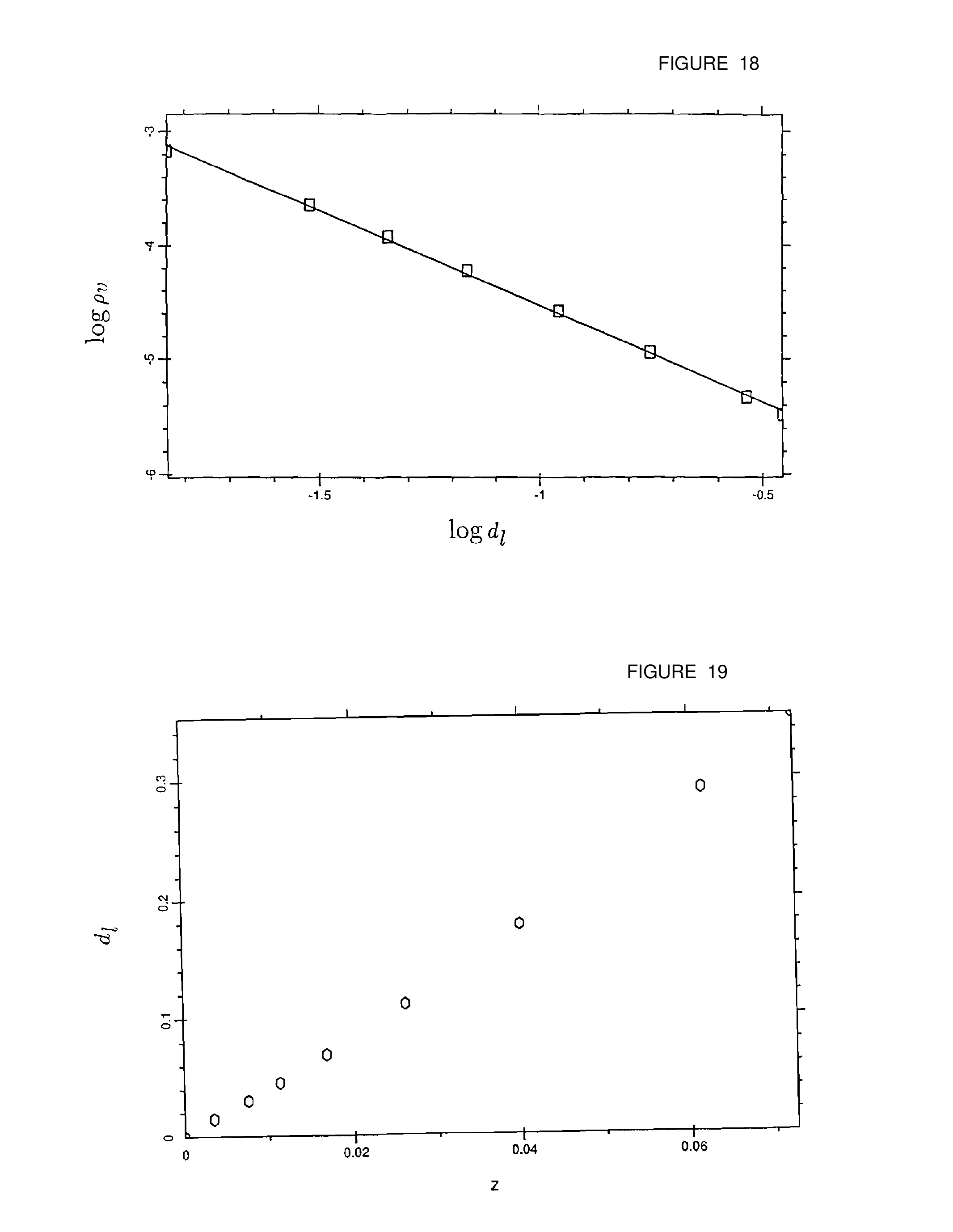}
\end{figure}
\begin{figure}
\includegraphics[height=18cm]{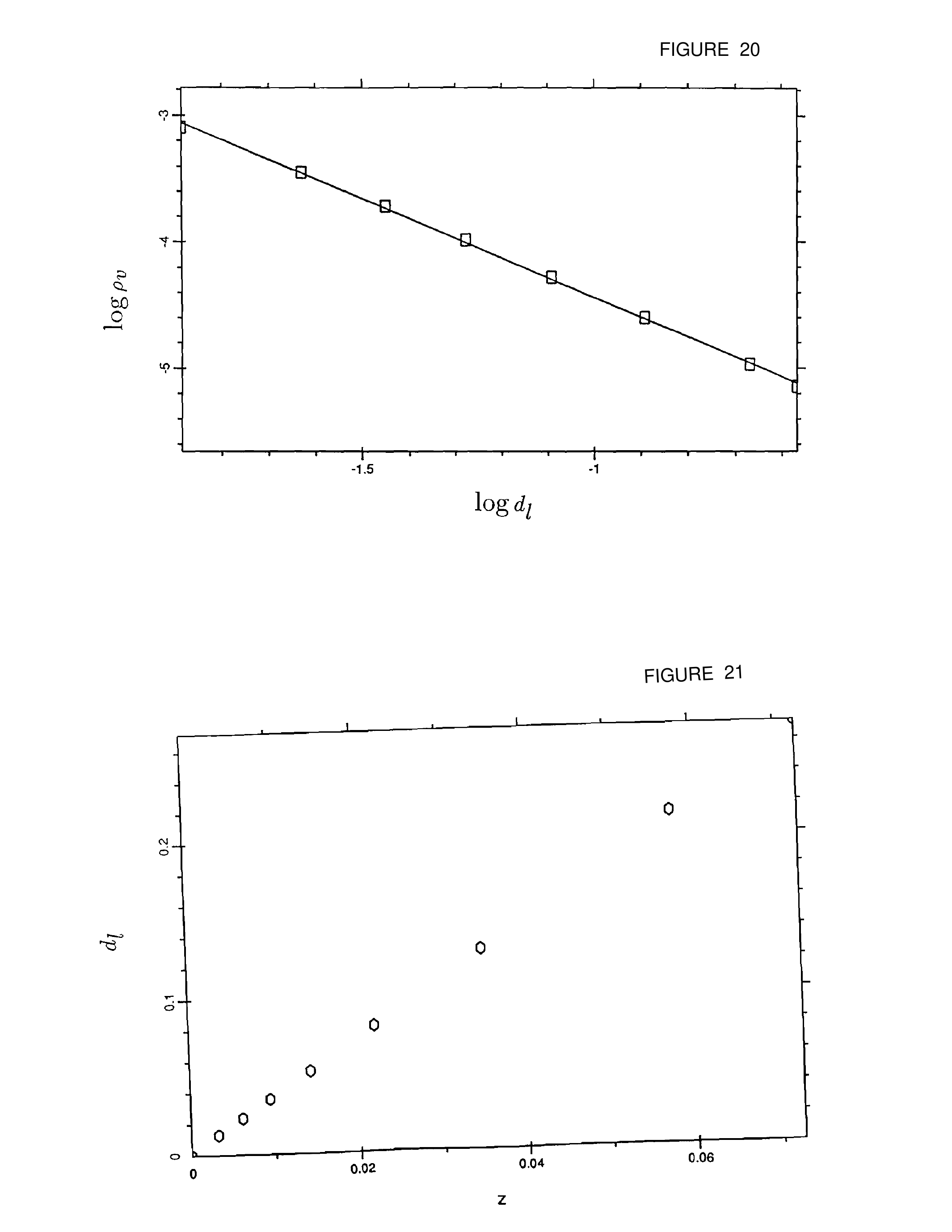}
\end{figure}
\begin{figure}
\includegraphics[height=18cm]{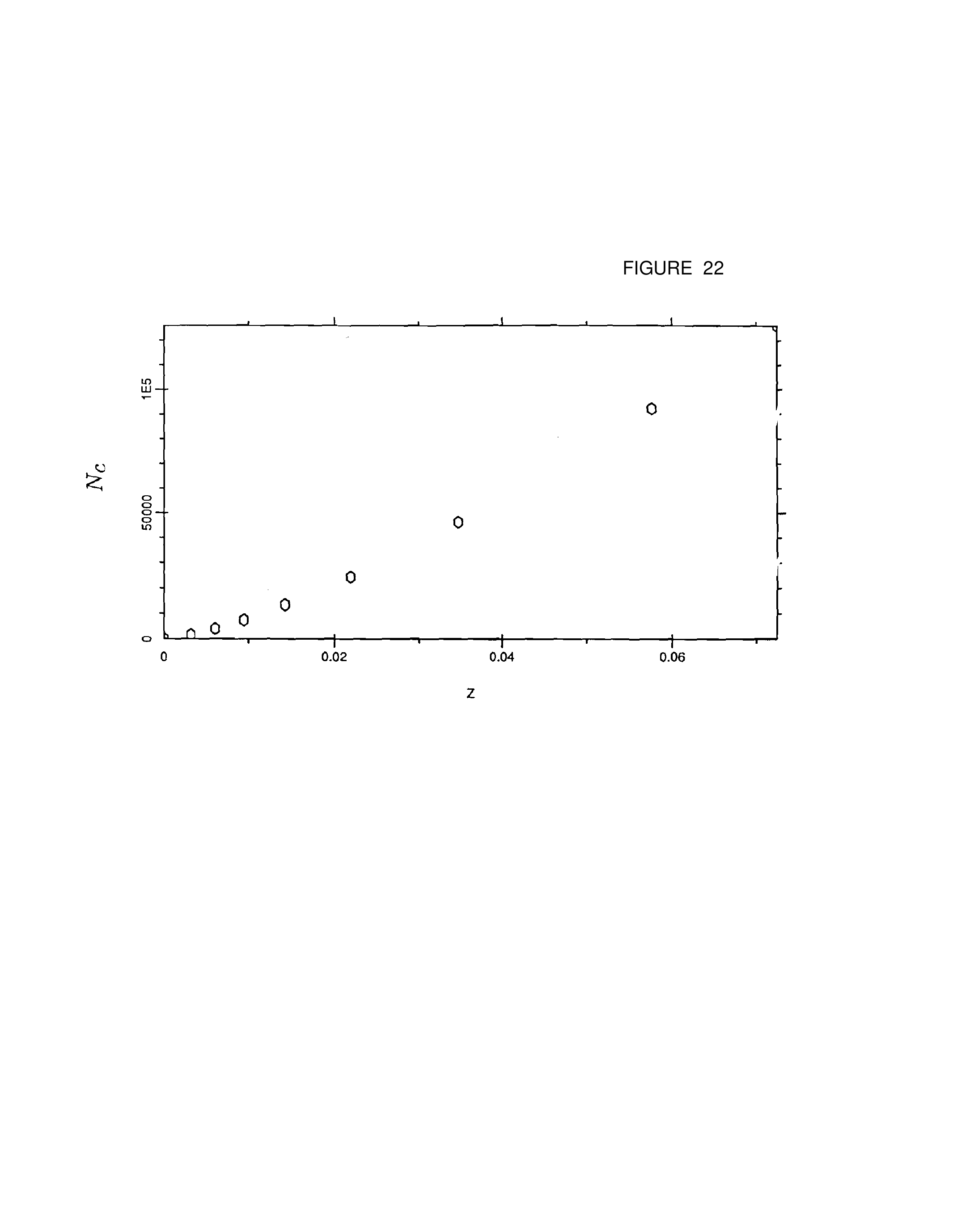}
\end{figure}
\end{document}